\documentclass[aps,prb,twocolumn,superscriptaddress,showpacs]{revtex4}

\usepackage{graphicx}
\usepackage{afterpage}

\begin{document}


\title{Magnetic and thermodynamic properties of cobalt doped iron pyrite:
Griffiths Phase in a magnetic semiconductor.}


\author{S. Guo}
\affiliation{Department of Physics and Astronomy, Louisiana State
University, Baton Rouge, Louisiana 70803 USA}

\author{D.P. Young}
\affiliation{Department of Physics and Astronomy, Louisiana State
University, Baton Rouge, Louisiana 70803 USA}

\author{R.T. Macaluso}
\affiliation{Department of Chemistry, Louisiana State
University, Baton Rouge, Louisiana 70803 USA}

\author{D.A. Browne}
\affiliation{Department of Physics and Astronomy, Louisiana State
University, Baton Rouge, Louisiana 70803 USA}

\author{N.L. Henderson}
\affiliation{Department of Physics and Astronomy, Louisiana State
University, Baton Rouge, Louisiana 70803 USA}

\author{J.Y. Chan}
\affiliation{Department of Chemistry, Louisiana State
University, Baton Rouge, Louisiana 70803 USA}

\author{L.L. Henry}
\affiliation{Department of Physics, Southern University, Baton Rouge,
Louisiana, 70813 USA}

\author{J.F. DiTusa}
\affiliation{Department of Physics and Astronomy, Louisiana State
University, Baton Rouge, Louisiana 70803 USA}


\date{\today}

\begin{abstract}
Doping of the band insulator FeS$_2$ with Co on the Fe site introduces
a small density of itinerant carriers and magnetic moments. The
lattice constant, AC and DC magnetic susceptibility, magnetization,
and specific heat have been measured over the $0\le x\le 0.085$ range
of Co concentration.  The variation of the AC susceptibility with
hydrostatic pressure has also been measured in a small number of our
samples. All of these quantities show systematic variation with $x$
including a paramagnetic to disordered ferromagnetic transition at
$x=0.007\pm 0.002$.  A detailed analysis of the changes with
temperature and magnetic field reveal small power law dependencies at
low temperatures for samples near the critical concentration for
magnetism, and just above the Curie temperature at higher $x$.  In
addition, the magnetic susceptibility and specific heat are
non-analytic around $H=0$ displaying an extraordinarily sharp field
dependence in this same temperature range. We interpret this behavior
as due to the formation of Griffiths phases that result from the
quenched disorder inherent in a doped semiconductor.

\end{abstract}

\pacs{75.50.Pp, 75.40.-s, 75.20.Hr}

\maketitle


\section{Introduction and Motivation}
The desire for ever increasing computational speeds has led to the
conception of spintronics technologies that make use of both the
charge and spin properties of electrical charge carriers in
solids\cite{wolf,vonmolnar}. Realization of these devices for useful
technologies requires the discovery and development of materials that
have easily controllable electronic and magnetic properties to the
same extent that semiconductors allow control over charge properties.
This demand has led to an enormous interest in magnetic semiconductors
to investigate if traditional semiconductors can be manipulated by
chemical substitution of transition metal elements, or other creative
means, into offering control over the spin properties of
electrons\cite{ohno1,ohno2,matsukura,reed,myers,salis}. Because so
many of the suggested devices rely on the production and detection of
spin polarized currents, magnetic semiconductors with Curie
temperatures above room temperature that are compatible with silicon
technology are highly desired. However, this goal has thus far not
been met and it has become increasingly apparent that to make progress,
fundamental knowledge and understanding of the electronic and magnetic
properties of magnetic semiconductors are necessary. Although some
headway in this direction has been made, the complexity of disordered
materials with strong interactions makes magnetic semiconductors
difficult to model and understand\cite{priour,galitski,schulthess}.

In a recent paper we presented the results of an exploration into the
magnetic and electronic properties of one such magnetic semiconducting
compound, Fe$_{1-x}$Co$_x$S$_2$ based on the diamagnetic insulating
parent compound iron pyrite (FeS$_2$)\cite{guo}. We found that the
insulator-to-metal transition that occurs for low concentrations of Co
doping was followed by a transition from a paramagnet to a disordered
ferromagnet first apparent for small dopant concentrations, $x$, at
very low temperatures. Increased Co doping led to an increased
transition temperature and to a more ordered ferromagnetic
phase\cite{Jarrett,bouchard,ramesha,wang,cheng}. The main conclusion
of our paper was that the quenched disorder associated with the
chemical substitution has important consequences on the formation of
the magnetic state at small $x$ and on the finite temperature phase
transitions at larger $x$. Unusual power law temperature, $T$, and
magnetic field, $H$, dependencies of the magnetic susceptibility,
magnetization, and specific heat were measured at low-$T$ for $x$ near
the critical concentration for nucleation of a magnetic phase, $x_c$,
as well as at $T$'s just above the Curie temperature, $T_c$ for
$x>x_c$.  These were attributed to large, rare disorder fluctuations
consisting of regions of local magnetic order which fluctuate as a
single large magnetic moment and that grow in size as $T \rightarrow 0$
or $T \rightarrow T_c$. Fluctuating rare region phenomena associated
with phase transitions in systems with quenched disorder are known
under the heading of Griffiths
phases\cite{Griffiths,vojtarev,vladrev}.

Theorists have found the subject of Griffiths phases fascinating since
Griffiths discovered the non-intuitive feature in his calculations
that, under certain conditions, exponentially rare regions in
disordered systems can dominate the thermodynamic response when in
proximity to a phase transition\cite{Griffiths,vojtarev}. This is in
contrast to many aspects of phase transitions where disorder can be
treated perturbatively or simply ignored. These ideas have drawn even
more interest as evidence for non-Fermi liquid behavior in disordered
systems has been discovered while explanations remain
tentative\cite{Stewart}. While it is clear that the disorder and
strong electron correlations are responsible for the breakdown of
Fermi liquid behavior in some cases, the microscopic origins remain
poorly understood\cite{vladrev}. In this context, the unusual power
law behavior associated with the formation of Griffiths phases have
been discussed as one mechanism for disorder driven non-Fermi liquid
behavior\cite{CastroNeto,castro2}. Experimental evidence that
Griffiths phases are observable in condensed matter systems has only
recently been reported for a small number of
systems\cite{ancona,salamon,deisenhofer,shimada,herrero}, and in our
previous publication we presented evidence for the importance of
Griffiths phase formation near the critical concentration for magnetic
ordering in the magnetic semiconducting system
Fe$_{1-x}$Co$_x$S$_2$. The purpose of the present paper is to present
a more thorough rendering of our magnetic susceptibility,
magnetization, and specific heat experiments and analysis, as well as
to present the results of some more recent AC magnetic susceptibility
measurements made at ambient and applied pressures of up to 7 kbar. In
a separate accompanying article we present the results of our Hall
effect and resistivity measurements of these same
materials\cite{guoprb2}.

We chose to investigate Co doped FeS$_2$ since FeS$_2$ is a simple
diamagnetic band insulator with the same pyrite crystal structure as
CoS$_2$, a itinerant ferromagnet\cite{Jarrett,bouchard,benoit}. The
two compounds had been shown earlier to be miscible such as to form
the substitutional series Fe$_{1-x}$Co$_x$S$_2$ without observation of
second phases or any phase segregation that occurs in many magnetic
semiconducting compounds. As such, it allowed the investigation of the
transition from diamagnetic insulator to a paramagnetic metal, then to
a ferromagnetic metal with Co substitution. In addition, band
structure calculations have predicted that at large $x$, $0.5\le x\le
0.95$, Fe$_{1-x}$Co$_x$S$_2$ is half metallic having a fully spin
polarized ground state\cite{ramesha,wang,mazin,cheng}

We begin our discussion by presenting the details of the experimental
techniques that we have used to investigate the extrinsic properties
of Fe$_{1-x}$Co$_x$S$_2$. This is followed by a presentation of the
magnetic susceptibility, magnetization, and specific heat over a wide
range of $x$ covering insulating, paramagnetic metallic, and
ferromagnetic metallic behaviors. In section \ref{Griffneartc} we
explore in detail the behavior of our crystals nearest $x_c$ at low
temperatures where the power-law in $T$ and $H$ behaviors are found.
This is followed by a discussion of the magnetic properties of
crystals with $x>x_c$ in proximity to $T_c$ displaying similar
behavior. We conclude the paper with a summary and discussion of our
results.

\section{Experimental Details}

Single crystals of Fe$_{1-x}$Co$_x$Si were synthesized from high
purity starting materials, including Fe powder (Alfa Aesar 99.998\%),
Co powder (Alfa Aesar 99.998\%), and sulfur (Alfa Aesar
99.999\%). These materials were sealed in a 16 mm diameter quartz tube
under vacuum. Iodine (Alfa Aesar 99.99+\%) was added to the tube as
the transport agent at a concentration level of 2 to 5 gm/cm$^3$. The
tubes were then placed in a small, single zone horizontal tube furnace
(Lindberg/Blue Model 55035) and heated to a constant
temperature. Insulating plugs were removed from the ends of the
furnace to create a natural temperature gradient of $\sim 50$
$^o$C. The charge was placed at the center of the furnace (hot zone),
and single crystals would begin to form at the cold end of the tube in
approximately 5 days, with a maximum yield by no later than 14
days. Good crystals were produced for temperatures in the range of 650
to 850 $^o$C. Crystals were etched in HCl to remove any remaining
flux.

\begin{figure}[htb]
  \includegraphics[angle=90,width=3.0in,bb=80 350 332 694,
clip]{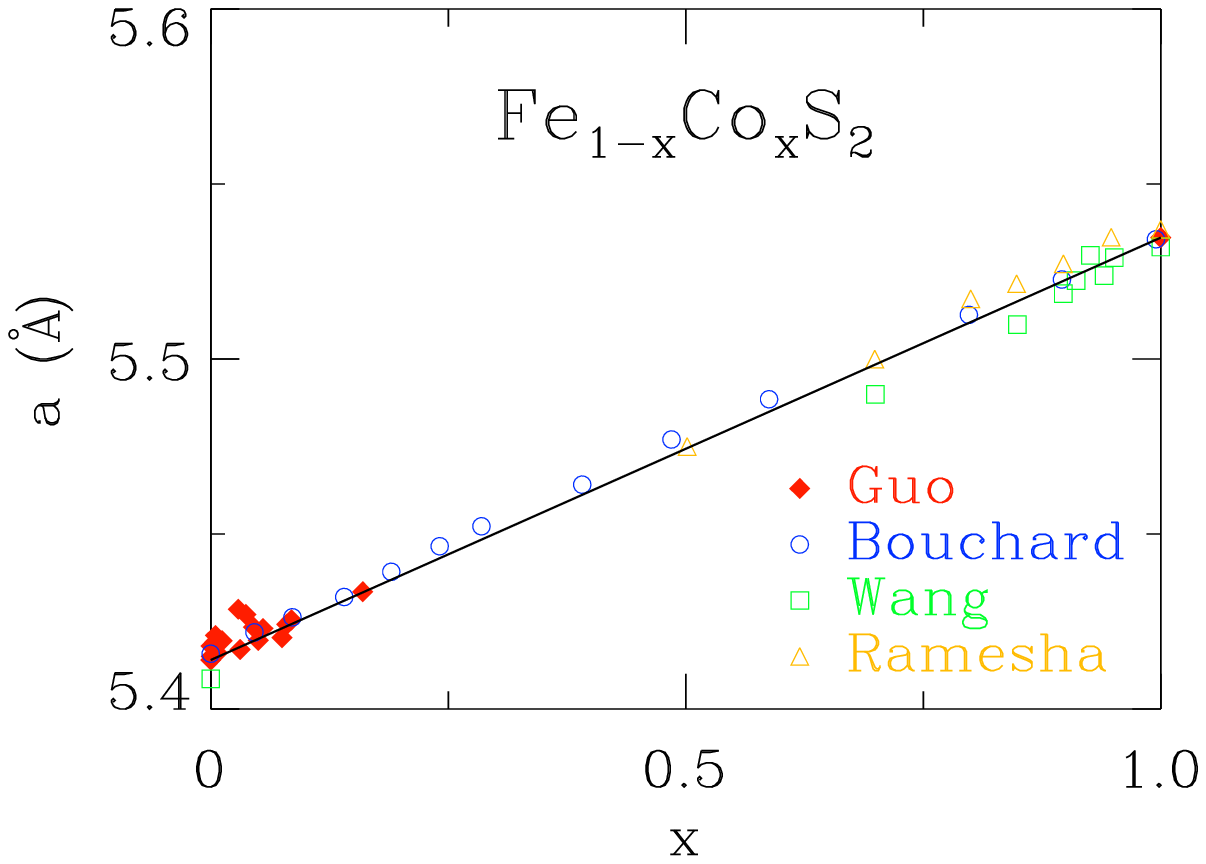}%
  \caption{\label{fig:latcon} (Color online) Lattice constants.
Lattice constant, $a$, for a wide range of cobalt concentration, $x$,
determined by single crystal X-ray diffraction for our samples
(labeled Guo) along with comparisons with those previously
determined{\protect{\cite{bouchard,wang,ramesha,cheng}}}. Black line
is a linear $x$ dependence of $a$ from FeS$_2$ through to CoS$_2$.  }
\end{figure}

Initial characterization included single crystal X-ray
diffraction. Single crystal fragments were glued to a glass fiber and
mounted on the goniometer of a Nonius Kappa CCD diffractometer
equipped with Mo-K$_{\alpha}$ radiation (wavelength $=
0.71073{\AA}$). All X-ray structure determinations are consistent with
a Pyrite crystal structure with no anomalous intensities that could be
interpreted as evidence for a second phase. The model of the structure
was refined to a level consistent with a fully solved crystal
structure. We observe no evidence for disorder, including that
associated with sulfur deficiencies, in the crystal structure of all
our crystals. The lattice constants, $a$, shown in
Fig.~\ref{fig:latcon} were determined by the X-ray diffraction
measurements.  As displayed in the figure, $a$ increases
systematically with Co concentration, $x$ in Fe$_{1-x}$Co$_x$S$_2$,
beyond the lattice constant for pure FeS$_2$, $a=0.54165$ nm, and is
comparable to the measurements of
Refs.[\onlinecite{bouchard,wang,ramesha,cheng}]. The increase in
lattice constant with $x$ is consistent with Vegard's law and the idea
that Co replaces Fe within the pyrite crystal
structure. Energy-dispersive X-ray microanalysis (EDX) on a JEOL
scanning electron microscope equipped with a Kevex Si(Li) detector was
performed to check the stoichiometry of our samples. These revealed
that the Co concentration was about $0.7\pm 0.1$ times the nominal Co
concentration of the starting materials, while the sum of the Co and
Fe densities was within error, $\pm 2$\%, of one half the sulfur
density.

Magnetization, $M$, and magnetic susceptibility (both AC and DC)
measurements were performed in a Quantum Design superconducting
quantum interference device (SQUID) magnetometer (MPMS) for
temperatures, $T \ge 1.8$ K and DC magnetic fields, $H$, of up to 5
T. AC susceptibility measurements were carried out over a range of
excitation fields from 0.1 to 1.5 Oe and excitation frequencies from 1
Hz to 1 kHz. In addition, AC magnetic susceptibility measurements were
performed in a dilution refrigerator above 50 mK at $H=0$, a frequency
of 1 kHz, and an excitation field of 1.5 Oe. The saturation
magnetization was determined by DC magnetization measurements at 1.8 K
and high fields (up to 5 T). These measurements were consistent with
our EDX measurements showing a magnetic moment density of $0.7\pm0.1$
times the nominal Co concentration. Because $M$ measurements are
easily performed and highly reproducible, we have subsequently used
the saturated magnetization to determine the Co concentration of our
samples. Thus, throughout this manuscript the stoichiometry of the
samples noted in the figures and text was determined in this
manner. The variations in the saturated magnetization for crystals
from the same growth batch was measured to be $\pm 10$\% of the
average value. Therefore, we report $x$ determined from measurements
of $M$ at high field for the crystals used in each of our
measurements. Wherever possible, the same crystal was employed for
several different measurement types, particularly for the measurement
of the AC susceptibility and specific heat below 1 K.

The magnetic susceptibility of several of our crystals was measured
with applied pressure ranging from ambient to 7 kbar in a
beryllium-copper pressure cell\cite{songthesis}.  Our pressure cell
was based on previously published
designs\cite{murata,koyama,kamishima}. The pressure cell was machined
from beryllium-copper alloy 25 since this material has a low magnetic
susceptibility and superior mechanical properties. After machining,
the cell was annealed in air at a temperature of 335 $^o$C for 2
hours. We designed our pressure cell to have a length of 12 cm to
reduce the background seen by the SQUID pickup coils originating from
the ends of the cell. The pistons are made from high purity Technox
3000 zirconia rods from Dynamic Ceramic. This resulted in a small,
diamagnetic, ($\sim -7\times10^{-6}$ emu at 50 Oe field) background
signal due to the lack of zirconia at the sample position. This
background depended slightly on the distance between the ends of the
two zirconia rods which, in turn, depended on the sample thickness,
increasing for larger separations. The sample was placed inside a
Teflon container which was sealed with a beryllium-copper cap and
retainer. The Teflon container is filled with silicon oil (Dow Corning
704) as the pressure transmission medium (see for example Ref.\
\onlinecite{carter} for more on silicon oil as a pressure transmission
medium). The pressure at the sample position was determined by placing
a small (0.5 mg) (V$_{0.99}$Ti$_{0.01}$)$_2$O$_3$ crystal in the
Teflon container for use as a manometer\cite{carter}. This material
makes an excellent manometer as it displays a readily measurable
metal-to-insulator transition which varies linearly in $T$ with
pressure for pressures of up to 10 kbar\cite{carter}.

The specific heat was measured using a standard thermal relaxation
method in a Quantum Design Physical Property Measurement System (PPMS)
above 2 K and a dilution refrigerator above 100 mK. For measurements
below 2 K single crystals were polished flat and mounted on a Sapphire
substrate with GE varnish. The opposite side of the 0.25 mm thick
Sapphire substrate contained a RuO resistance thermometer which was
notched to provide a thermometer-heater pair in good thermal contact
with the sample. The thermometer was cycled from 300 K to 4.2 K many
times to ensure repeatability of the resistance at low
temperatures\cite{goodrich}. The thermometer was calibrated by
comparison to a Lakeshore calibrated germanium resistance thermometer.
Four resistive wires provided electrical and thermal contact of the
thermometer and heater to the mixing chamber of the dilution
refrigerator.  We carefully checked at each measurement temperature
that the time constant for thermalization of all elements on the
Sapphire substrate was much smaller than the thermal time constant for
thermalization to the cryostat. Specific heat measurements were
carried out in magnetic fields between 0 and 6 T with all thermometers
calibrated at each magnetic field. The specific heat of the addenda
was measured separately and carefully subtracted from the data.

\section{Overview of Experimental Results}
\subsection{DC susceptibility and Magnetization Measurements
\label{DCsusceptibilityandMagnetizationMeasurements}}
\begin{figure}[htb]
  \includegraphics[angle=90,width=3.0in,bb=75 300 560 710,
clip]{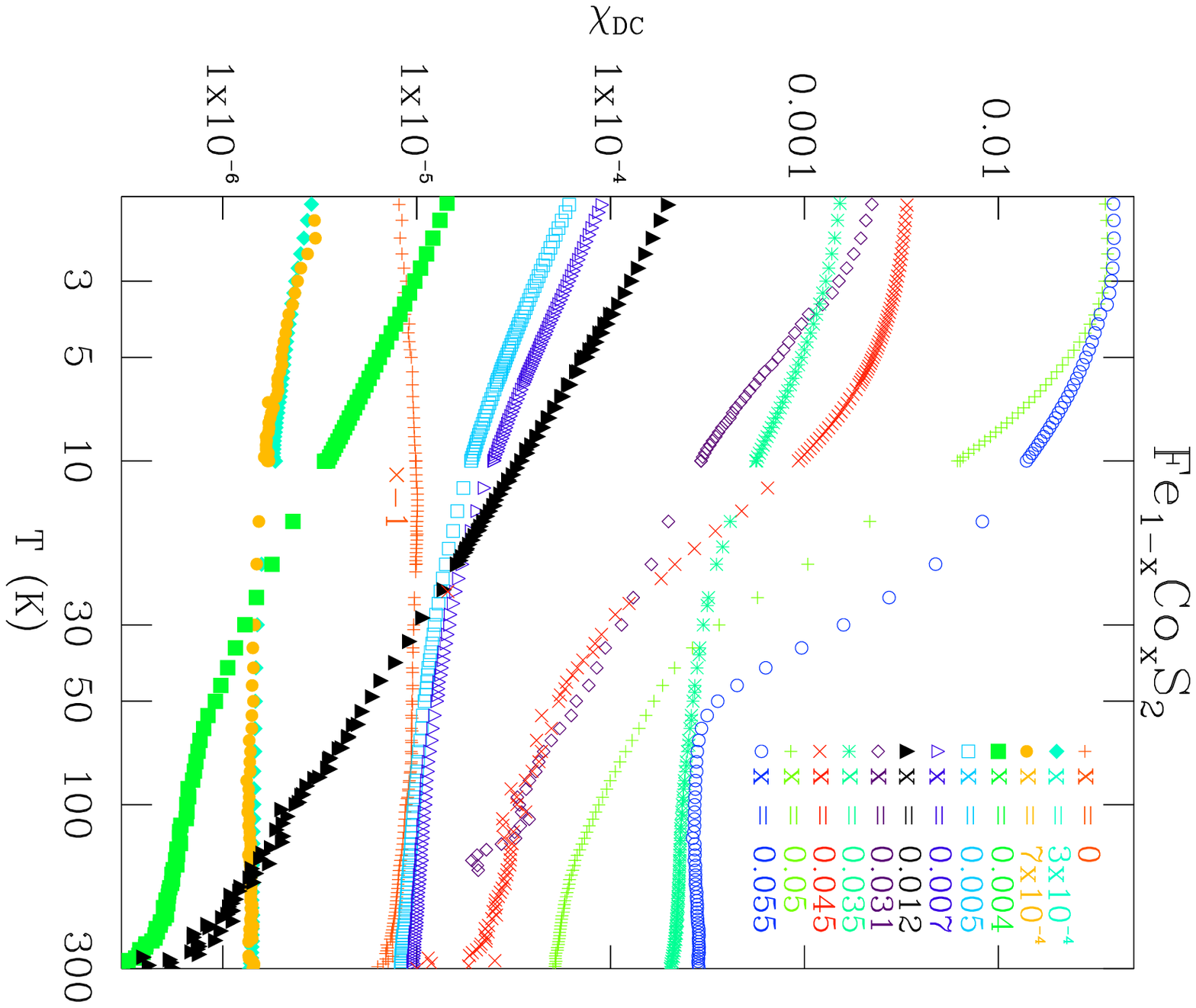}%
  \caption{\label{fig:chidc} (Color online) DC magnetic
susceptibility. Temperature, $T$, and cobalt concentration, $x$,
dependence of the DC magnetic susceptibility for several
representative crystals. Note that the data for the $x=0$, nominally
pure sample, is diamagnetic and thus has been multiplied by $-1$ for
display purposes. Magnetic fields applied include 10 kOe for the
$x=3\times10^{-4}$, $7\times10^{-4}$, and 0.004 crystals; 2 kOe for
the $x=0$, 0.012, and 0.035 crystals; 1 kOe for the $x=0.005$, 0.007,
and 0.045 crystals; and 50 Oe for the $x=0.031$, 0.05, and 0.055
crystals. Measurement fields were chosen so that noise levels were
acceptable given the size of the susceptibility signal for each
individual crystal.}
\end{figure}

In order to establish the magnetic ground states and critical
temperatures of the Fe$_{1-x}$Co$_x$S$_2$ chemical substitution series
we have measured the DC magnetic susceptibility, $\chi_{DC}$, and
magnetization, $M$, of our crystals. In Fig.~\ref{fig:chidc} we plot
$\chi_{DC}$ of a dozen representative samples with $x$ ranging from
nominally pure FeS$_2$ to $x=0.055$. Our $x=0$ crystal has a
diamagnetic susceptibility that changes little as a function of
temperature. The small size of the impurity related Curie-Weiss
contribution to the susceptibility of this crystal observable below 20
K, an increase of $10^{-6}$ from 20 K down to 2 K, is consistent with
an impurity concentration of $3\times10^{18}$ cm$^{-3}$. Not only does
this establish the quality of our pyrite crystals, it is compelling
because it reaffirms that the parent compound is non-magnetic despite
being 1/3 iron\cite{Jarrett,bouchard}. The low temperature $\chi_{DC}$
systematically increases with $x$ by over 4 orders of magnitude for
our $x=3\times10^{-4}$ to our $x=0.055$ samples. At the same time, the
$T$-dependence changes from being almost Curie-like at low-$x$, to
displaying a broad, almost constant region of $\chi_{DC}$ below 5 K
for $x> 0.035$. At higher temperatures $\chi_{DC}$ is less systematic
in $x$ with some crystals displaying almost constant $\chi_{DC}$ above
100 K suggestive of a significant Pauli susceptibility, while others
show a more $T$-dependent $\chi_{DC}$ to higher temperatures crossing
through $\chi_{DC}$ of samples with smaller $x$. Because the Pauli
susceptibility is proportional to the conduction electron density of
states, we interpret the variation of our measured $\chi_{DC}$ above
100 K as revealing a carrier density and, perhaps, a mass that is
highly sensitive to the disorder. Because of this change from a
systematic low-temperature $\chi_{DC}$ to a less systematic high-$T$
$\chi_{DC}$, we conclude that in the two temperature ranges
$\chi_{DC}$ is dominated by two different electron populations; a more
local population of electron moments dominates at low-$T$ while the
more itinerant electrons dominate $\chi_{DC}$ at $T>100$ K.

\begin{figure}[htb]
  \includegraphics[angle=90,width=3.0in,bb=75 300 560
  725,clip]{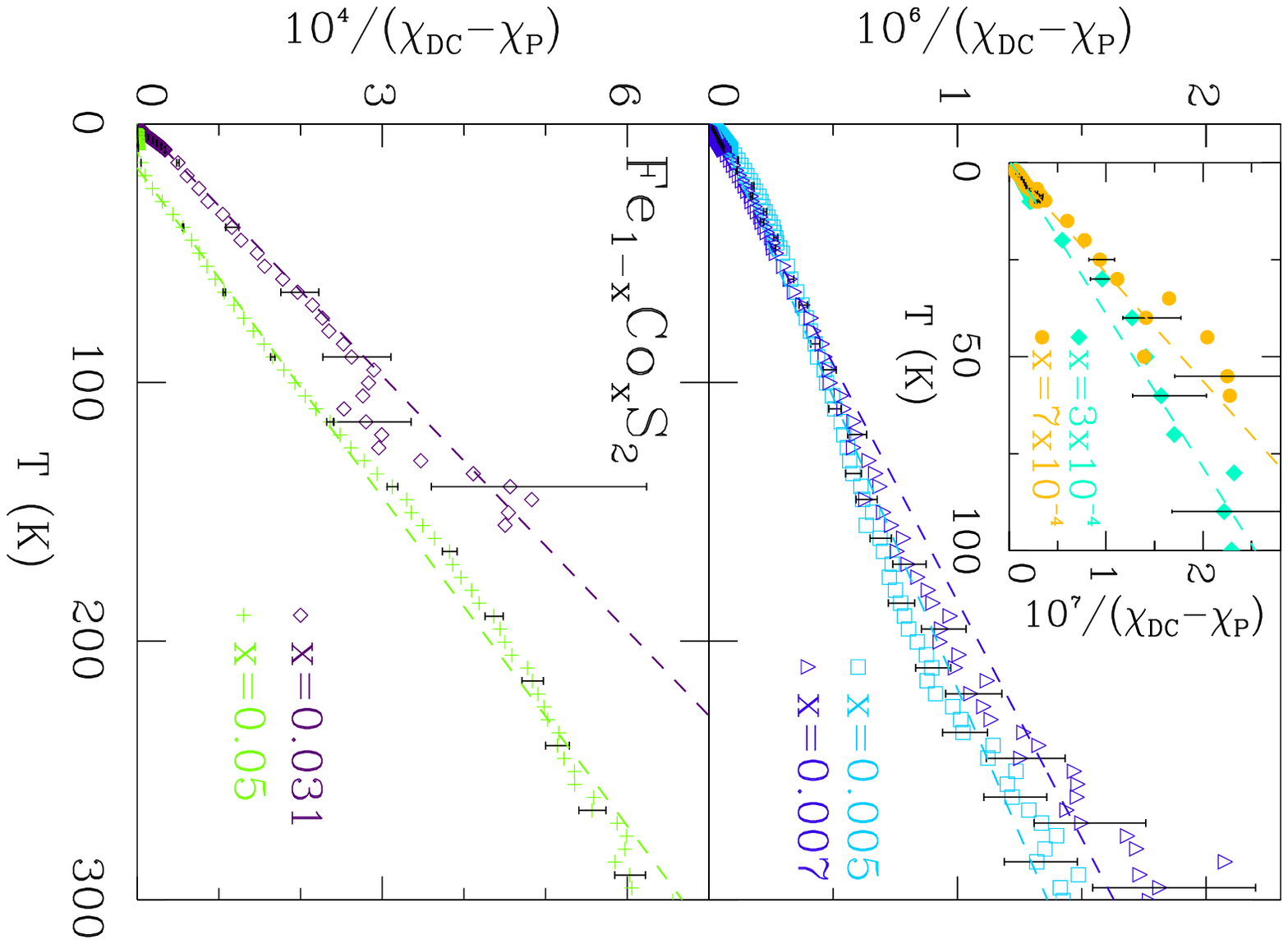}%
  \caption{\label{fig:cwpl} (Color online) Inverse DC magnetic
susceptibility. a) and b) Temperature, $T$, and cobalt concentration,
$x$, dependence of the inverse DC magnetic susceptibility,
$1/\chi_{DC}$, after subtraction of the temperature independent
contribution, for several representative samples at the same magnetic
fields as in Fig.~{\protect\ref{fig:chidc}}. The constant
susceptibility is attributed to the Pauli susceptibility, $\chi_P$ of
our crystals. Inset to frame a) displays $1/(\chi_{DC} - \chi_P)$
below 100 K for our $x=3\times10^{-4}$ and $x=7\times10^{-4}$ where
the signal to noise level is such that Curie behavior is apparent in
an inverse susceptibility plot. Error bars are plotted only for every
5th (3rd in inset) data point for display purposes.}
\end{figure}

\begin{figure}[htb]
  \includegraphics[angle=90,width=3.0in,bb=75 350 535
  715,clip]{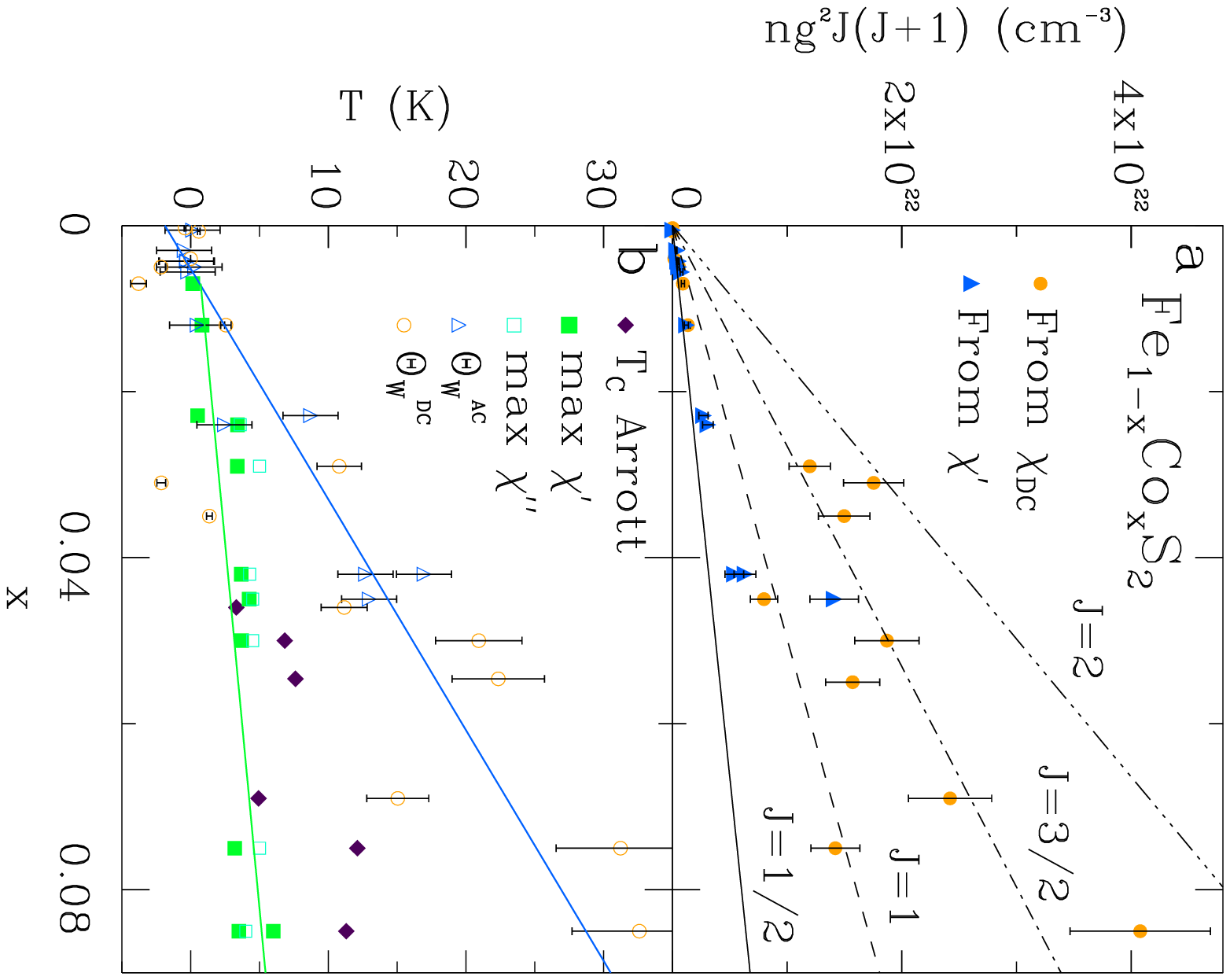}%
  \caption{\label{fig:tcpl} (Color online) Curie constant, Curie
temperature and Weiss temperature. a) The Curie constant,
$ng^2J(J+1)$, determined from fits of a Curie-Weiss form added to a
temperature independent Pauli contribution, $\chi_P$, $\chi =
n(g\mu_B)^2 J J(+1) / 3 k_BT + \chi_P$, to the DC and AC magnetic
susceptibility. Lines indicate values of the Curie constant for
various $J$'s when $n$ is set equal to the cobalt concentration of the
crystals and $g$ is set equal to 2. b) The Curie temperature, $T_C$ as
defined by the temperature of the maximum in the real part of the AC
susceptibility, max $\chi'$, and through a mean field Arrott analysis,
$T_C$ Arrott (see text for details). The temperature of the maximum in
the imaginary part of the AC susceptibility, max $\chi''$ is included
in the figure for comparison to max $\chi'$. Weiss temperatures as
determined from the temperature dependence of the AC, $\Theta_W^{AC}$,
and DC, $\Theta_W^{DC}$, susceptibility are also plotted. Lines are fits
of a linear max $\chi'(x)$ and $\Theta_W^{AC}(x)$ dependence to the data. }
\end{figure}

To gain insight into the character of the magnetic moments responsible
for the low-$T$ upturn in $\chi_{DC}$ we have plotted in
Fig.~\ref{fig:cwpl} the inverse DC magnetic susceptibility after
subtraction of the temperature independent Pauli susceptibility,
$\chi_P$, $1/(\chi_{DC} - \chi_P)$ as a function of $T$. This is
standard procedure for comparing the susceptibility data to a
Curie-Weiss form, $\chi = CC/(T-\Theta_W^{DC})$ where $CC$ is the
Curie constant and $\Theta_W^{DC}$ is the Weiss temperature.  Although
some nonlinearity is apparent in the data, we have quantified the
Curie Weiss behavior by fitting a linear $T$-dependence to the data as
shown in the figure. We interpret the intercept of the fits with the
$T$-axis as $\Theta_W^{DC}$ and the slope as $1/CC$ where $CC=n
(g\mu_B)^2 J(J+1)/ 3k_B$ with $n$ the density of magnetic moments of
size $J$, $g$ the gyromagnetic factor, $k_B$ Boltzmann's constant, and
$\mu_B$ the Bohr magneton. Both $\Theta_W^{DC}$ and $CC$ are observed
to increase systematically with $x$ as can be seen in
Fig.~\ref{fig:tcpl} where we plot the best fit values of the quantity
$ng^2J(J+1) = 3CC k_B / \mu_B^2$ and $\Theta_W^{DC}$. It is also
interesting to note that for small $x$, $x < 0.01$, $\Theta_W^{DC}$ is
less than zero revealing a small antiferromagnetic interaction of the
moments for small Co doping. This is perhaps not expected since at
only slightly larger Co concentrations a disordered ferromagnetic
ground state results.

We note that the Curie constant increases with $x$ beyond what would
be expected for paramagnetic spin-1/2 magnetic moments for all samples
measured. If we assume that the density of magnetic moments is equal
to the Co density, and that $g=2$ we would conclude that the magnetic
moment increases from just above $J=0.5$ to between $J=1$ and $J=2$
for $x\ge 0.03$. This is the first indication that the magnetic moments
likely form strongly interacting clusters over the temperature scale
where the Curie constants were determined. The measured Curie
constant, in this case, most likely reflects the fluctuations of a
smaller density of much larger composite magnetic moments.

\begin{figure}[htb]
  \includegraphics[angle=90,width=3.0in,bb=75 300 560
  710,clip]{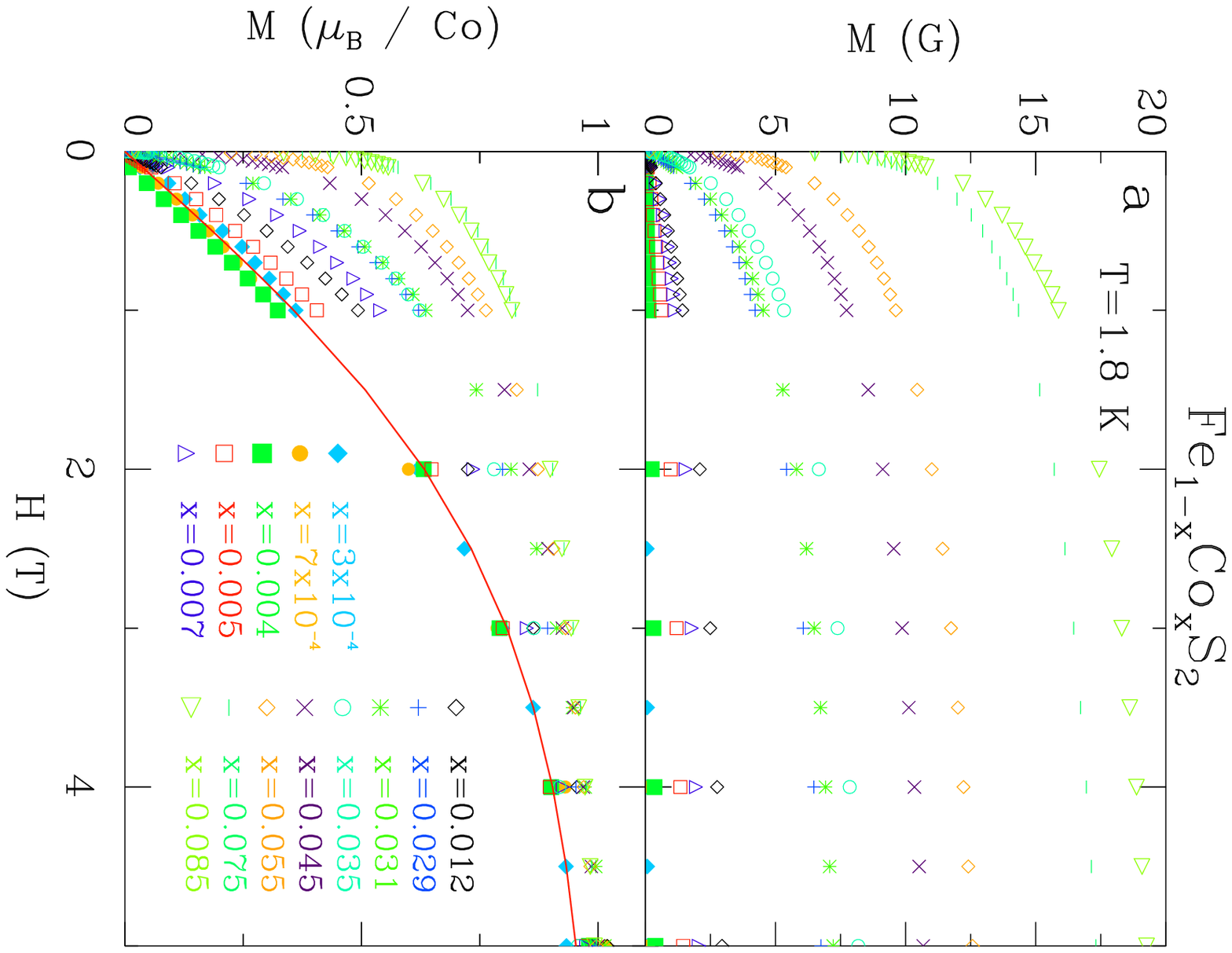}%
  \caption{\label{fig:magpl} (Color online) Magnetization. a) The
magnetization vs external magnetic field $H$ at 1.8 K for several of
our crystals with Co concentrations identified in frame b. b) The
magnetization normalized so that the high field values approach
1. This normalization was used along with the EDX data to determine
$x$. The red line is a Brillouin function for $J=1/2$ and a
gyromagnetic ratio of 2. }
\end{figure}

The magnetic properties of our samples was further explored by
measuring $M$ at 1.8 K in fields of up to 5 T. Fig.~\ref{fig:magpl}a
displays the enormous increase in $M$ with $x$ for a range of $x$
between $3\times 10^{-4}$ and 0.085. In Fig.~ \ref{fig:magpl} b we
display $M$, in $\mu_B$ per Co dopant, to demonstrate the
normalization used to determine the Co density of our crystals. The
form of the magnetization can be seen to evolve from that of a
Brillouin function, $B_J$, at small $x$ to that of a ferromagnetically
ordered material at larger $x$.

\begin{figure}[htb]
  \includegraphics[angle=90,width=3.5in,bb=60 320 550
  720,clip]{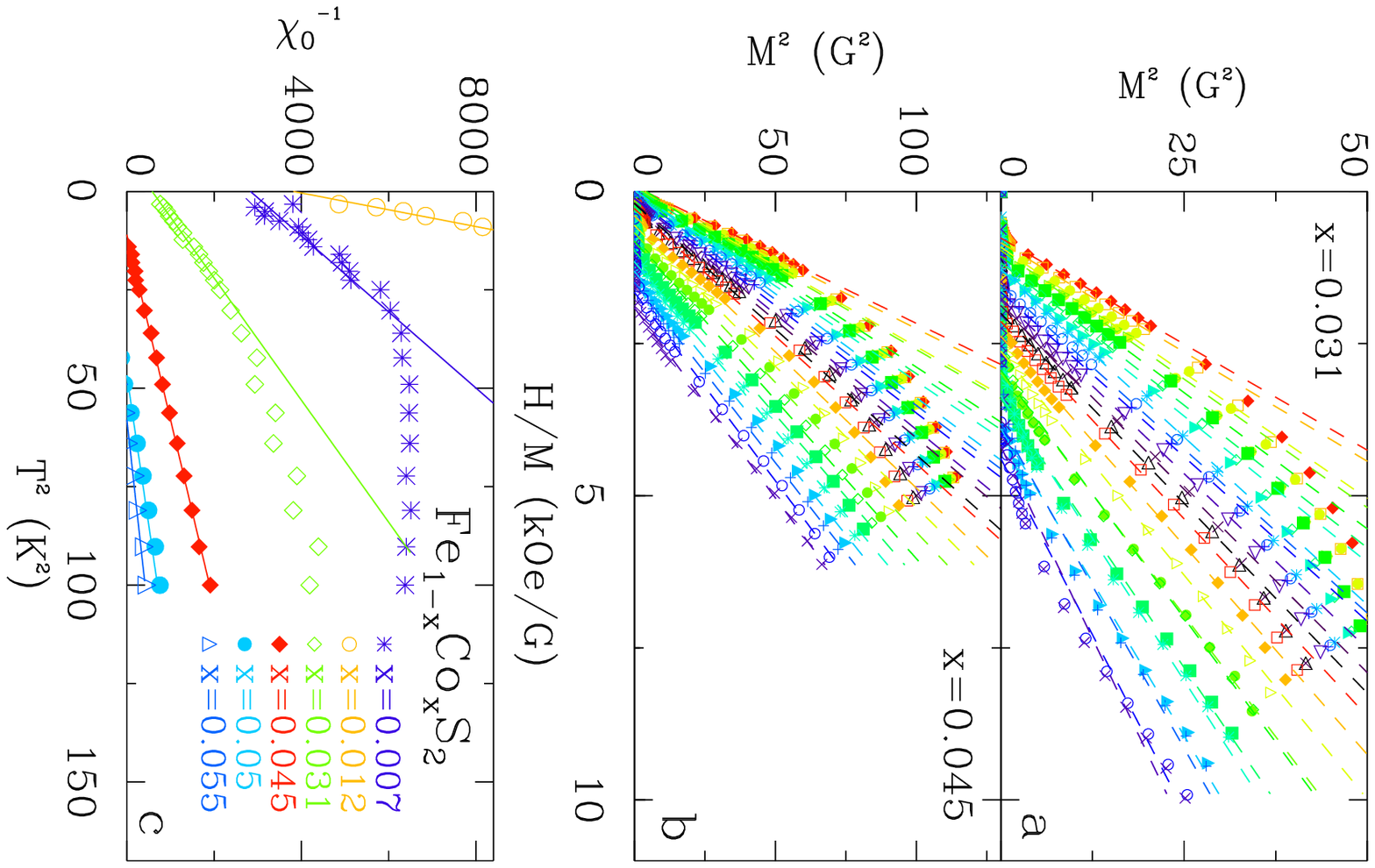}%
  \caption{\label{fig:arrott} (Color online) Arrott plots. Standard
mean field analysis for determining the Curie temperature, $T_c$, of
ferromagnetic materials. In a) and b) we plot the square of the
magnetization, $M$, vs.\ the external field, $H$, divided by the
$M$. $T_c$ is the temperature where the extrapolation of the data to
$H=0$ goes through 0. a) Arrott plot for an $x=0.031$ crystal. Data
are at temperatures of 1.8 (red filled diamonds), 2.0 (orange
squares), 2.25 (yellow bullets), 2.5 (yellow-green open triangles),
2.75 (green filled squares), 3.0 (green open diamonds), 3.25 (light
blue filled triangles), 3.5 (blue asterisks), 3.75 (blue-violet
circles), 4.0 (purple pluses), 4.25 (violet open right pointing
triangles), 4.5 (dark violet X's), 4.75 (black left open left pointing
triangles), 5.0 (red open squares), 5.5 (orange filled diamonds), 6.0
(yellow open triangles), 6.5 (yellow green bullets), 7.0 (green open
diamonds), 7.5 (blue-green filled squares), 8.0 (light blue
asterisks), 8.5 (blue filled triangles), 9.0 (blue pluses), 9.5
(blue-violet circles), and 10 K (violet X's). Dashed lines are linear
fits to data between 1 and 15 kOe.  b) Arrott plot for an $x=$0.031
crystal. Temperatures and symbols same as in frame a. Dashed lines are
linear fits to data between 1 and 20 kOe. c) Plot of $H/M$ axis
intercepts, $\chi^{-1}_0$, of the linear fits to data in frames a and
b vs $T^2$ for several crystals identified in the figure.  The
intercepts of these data with the $T^2$ axis is interpreted as the
mean-field $T_c$. }
\end{figure}

The temperature and field dependence of $M$, such as that shown in
Fig.~\ref{fig:magpl}, is commonly compared to a mean field form to
determine $T_c$ of magnetic materials. In Fig.~\ref{fig:arrott} we
display this comparison of $M(H,T)$ for two representative
samples. The mean field form for $M$ is typically written as $M(H) =
ngJ\mu_B\: B_J(gJ\mu_B\,(H+\eta M) / k_BT)$ with $\eta$ the constant
parameterizing the strength of the molecular mean field. From this
starting point an expansion for $H/M$ can be written as
$H/M=(1/\chi_0) + aM^2 +bM^4 + ...$ where $\chi_0$ is the initial
susceptibility ($\chi(H=0)$). Thus, a plot of $M^2$ vs.\ $H/M$ should
be a straight line for a range of fields such that the higher terms
can be ignored\cite{Arrott}. This is commonly referred to as an Arrott
plot and is demonstrated by Fig.~\ref{fig:arrott} frames a and b
showing the results of such an analysis for two samples near the
critical Co concentration for ferromagnetism. It is clear that the
region of linearity of $M^2$ in $H/M$ is limited, particularly at low
temperatures.  Frame c of the figure displays a compilation of
$1/\chi_0$ values determined from the linear fits, such as those in
frames a and b. We have plotted the values of $1/\chi_0$ vs.\ $T^2$
since the Stoner-Wohlfarth theory of itinerant magnetism predicts a
$T^2$ dependence of $1/\chi_0$ near the transition temperature. Linear
fits to the low-$T$ data are used to determine the $T^2$ intercept and
this is interpreted as the mean-field $T_c$. The values for $T_c$
determined in this way are plotted in Fig.~\ref{fig:tcpl} b for
comparison to $\Theta_W^{DC}$. Modified Arrott analysis's having
different dependencies of $H/M$ on $M$ were also performed without
significant changes to the critical temperatures determined from the
above analysis\cite{Arrott}. The conclusion from this mean-field
analysis is that the Curie temperature is less than $1/2\Theta_W^{DC}$
for these samples reflecting the importance of disorder on the
formation of the magnetic state and that there is a critical Co
concentration for ferromagnetism near $x=0.03$. However, a simple mean
field analysis is likely to be inaccurate when $T_c$ is close to zero.

\clearpage

\subsection{AC Susceptibility Measurements 
\label{ACSusceptibilityMeasurements}}
\begin{figure}[htb]
  \includegraphics[angle=90,width=3.0in,bb=80 375 530
  710,clip]{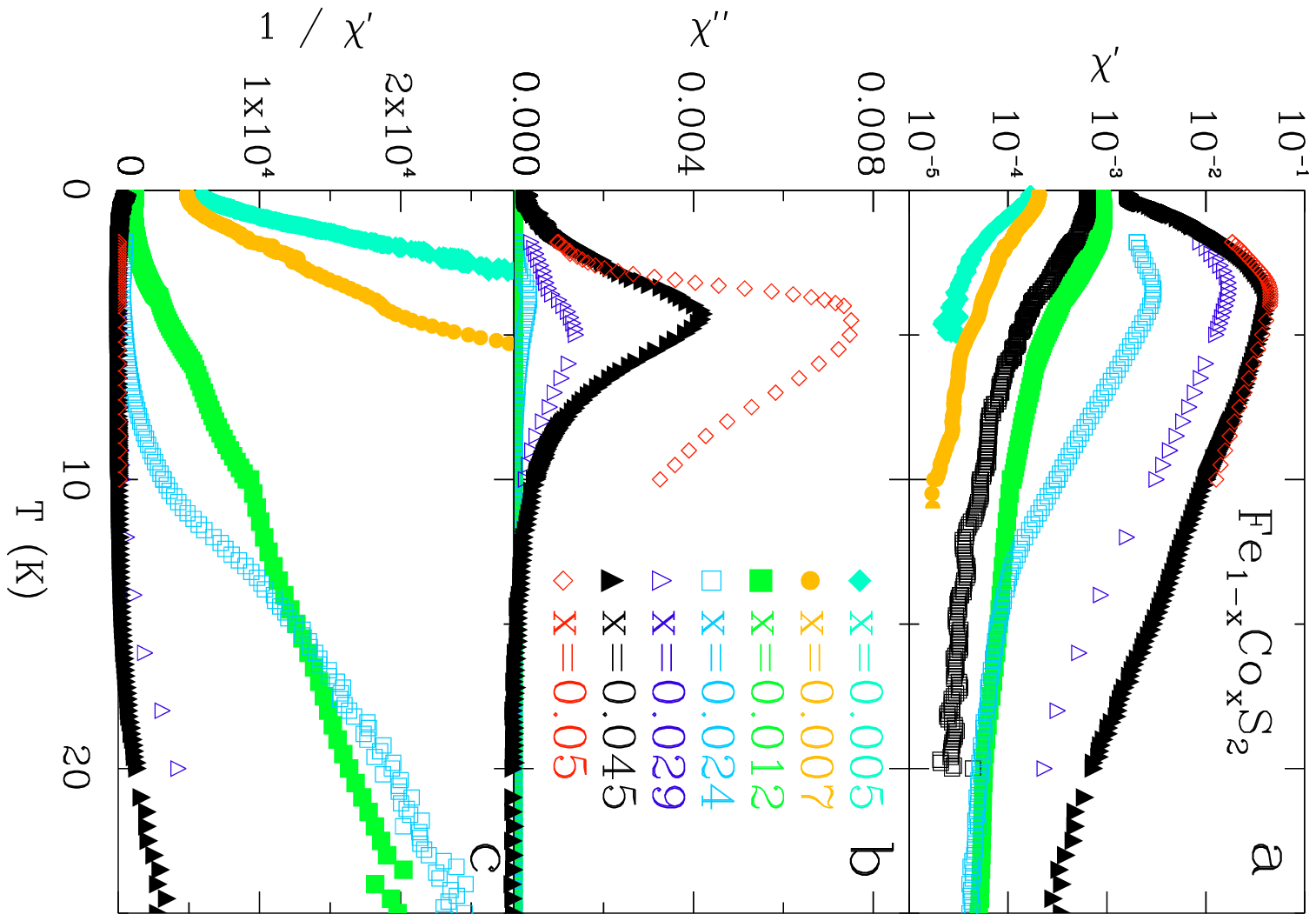}%
  \caption{\label{fig:chipl} (Color online) AC susceptibility. a) The
real part of the AC magnetic susceptibility, $\chi'$ of several of our
crystals with Co concentrations identified in frame b. The AC
excitation field was 1.5 Oe and the measurement frequency was 1.0 kHz
for the $x=$0.005, 0.007, 0.012, 0.024, and 0.045 crystals, 1.0 Oe and
10 Hz for the $x$=0.029 crystal, and 1.0 Oe and 100 Hz for the
$x$=0.05 crystal.  b) Imaginary part of the AC susceptibility,
$\chi''$, for the same crystals in frame a. Co concentrations are
identified in the figure. c) $1 / \chi'$ demonstrating Curie
Weiss-like behavior. Data and symbols same as in frame a and b. }
\end{figure}

Although the DC susceptibility and magnetization give us a good
initial indication of the formation of a ferromagnetic phase with the
substitution of Co for Fe in FeS$_2$, the very subtle changes that
occur near $x_c$ are best explored at very low magnetic fields where
good quality DC magnetization signals are difficult to
obtain. Therefore, we have made very careful AC susceptibility
measurements of our crystals using excitation fields of 1.5 Oe or less
as shown in Fig.~\ref{fig:chipl}. In agreement with our DC results,
the AC susceptibility shows a dramatic increase with
$x$. However,there are two important differences between our DC
and AC susceptibility apparent in the figures. First, the real part of
the AC susceptibility, $\chi'$, is noticeably larger than $\chi_{DC}$
for all $x>0$ crystals measured. Second, distinct maxima appear in
both $\chi'$ and, at slightly higher $T$'s, the imaginary part part of
the AC susceptibility, $\chi''$, as can be seen in frames a and b. As
is common practice, we preliminarily interpret the temperature of the
maxima in $\chi'$ as the critical temperature, $T_c$, for the
development of a magnetic, ferromagnetic or spin glass, phase. In
Fig.~\ref{fig:tcpl}b we plot $T_c$ determined in this manner (max
$\chi'$) along with the temperature of the maximum in $\chi''$ (max
$\chi''$) for comparison to our analysis of $\chi_{DC}$ and $M(H)$. In
contrast to our $\chi_{DC}$ results, our very low-$T$ measurements
(down to 50 mK) reveal a peak in $\chi'$ for all crystals with $x \ge
0.007$ indicating that $x_c = 0.007 \pm 0.002$. The contrasting results
in our AC and DC magnetic susceptibility indicate that the magnetic
state of these crystals are extraordinarily sensitive to magnetic
field as we explicitly demonstrate in section \ref{finitetc} below.

In frame c of Fig.~\ref{fig:chipl} $1/\chi'$ is plotted against
temperature to display the Curie-Weiss-like behavior of $\chi'$. Here,
the general Curie-Weiss-like trend is displayed although significant
nonlinearity and structure are also apparent in these data. Linear fits
of these data were used to determine the Weiss temperatures,
$\Theta_W^{AC}$, and Curie constants displayed in
Fig.~\ref{fig:tcpl}. There appears to be good compatibility of both
$\Theta_W^{AC}$ with the values determined from the DC values,
however, the Curie constant appears generally smaller for our AC
analysis than for our DC values for $x>0.02$. The Curie constants
determined from both methods indicate a fluctuating magnetic moment of
$J>1/2$ per Co dopant indicative of magnetic cluster
formation. Despite the general linear behavior of $1/\chi'$ well above
$T_c$, there is a noticeable departure from this behavior at
$T<4\Theta_W^{AC}$ that can best be seen for our $x=0.024$ sample in
frame c of Fig.~\ref{fig:chipl}. Fits of $1/\chi'(T)$ for $T>$ 15 K
indicate $\Theta_W^{AC} = 3$ K, however, at $T< 15$ K $1/\chi'$ tends
toward much smaller values. This tendency for $\chi'$ to increase at a
rate faster than the Curie-Weiss behavior indicates local
ferromagnetism and further supports our observation that magnetic
clusters form at temperatures above $T_c$. A more complete description
of this behavior is presented in section \ref{finitetc} of this paper.

\begin{figure}[htb]
  \includegraphics[angle=90,width=3.0in,bb=50 240 560
  700,clip]{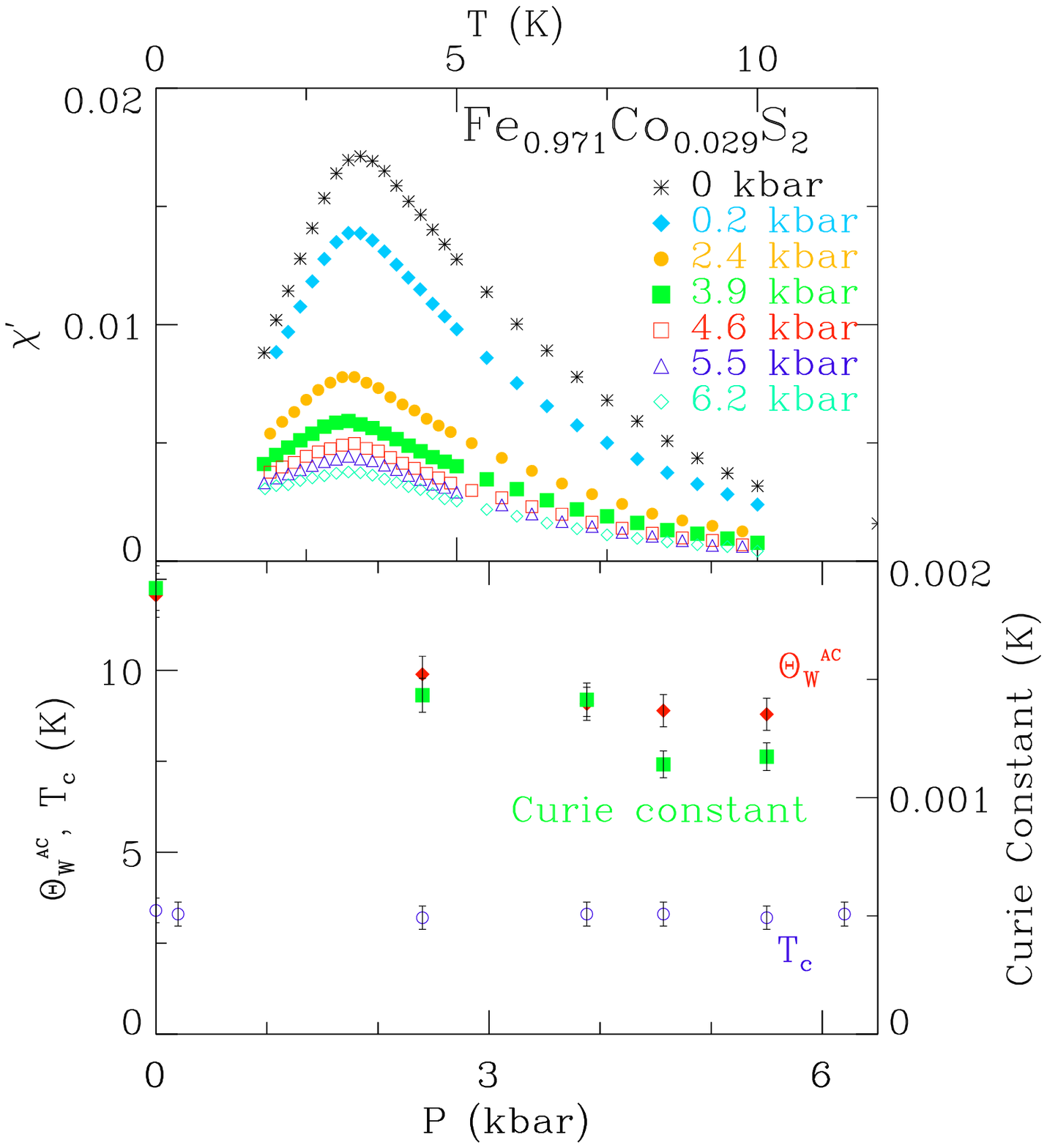}%
  \caption{\label{fig:chiprpl} (Color online) Pressure dependence of
the AC magnetic susceptibility. a) The pressure, $P$ and temperature,
$T$, dependence of the real part of the AC magnetic susceptibility,
$\chi'$, for an $x=0.029$ crystal at zero DC magnetic field and at
pressures identified in the figure.  Data taken with an excitation
field of 1 Oe at a frequency of 1 Hz. b) Curie temperature, $T_c$, as
defined by the temperature of the peak susceptibility in frame a, of
our $x=0.029$ crystal as a function of $P$. Also shown are the Weiss
temperature, $\Theta_W$, and Curie constant derived from fits of the
Curie-Weiss form to the real part of the AC susceptibility shown in
frame a.  Note: $\Theta_W$ and the Curie constant were determined only
for pressures where sufficient $T>15$ K $\chi'$ data (not shown) was
available. }
\end{figure}

The effect of hydrostatic pressure, $P$, on the magnetization of our
crystals is demonstrated in Fig.~\ref{fig:chiprpl}. While the
magnitude of $\chi'$ is depressed by the application of pressures of
order of a kbar, there is little change to the temperature of the peak
in $\chi'$. The changes that occur with $P$ are demonstrated in frame
b of the figure where the temperature of the peak in $\chi'$, taken as
our definition of $T_C$, is plotted along with the results of fitting
a Curie Weiss form to $\chi'(T)$ above 15 K. Both the Curie constant
and the Weiss temperature are reduced as $P$ is applied showing that
$P$ reduces both the average size of magnetic clusters as well as the
average interaction energies between magnetic moments evident in the
$P=0$ measurements. The Curie constant shown in Fig.~\ref{fig:chiprpl}
is consistent with a reduction of the density of $J=1/2$ magnetic
moments from $3\times10^{21}$ to $1.7\times10^{21}$ cm$^{-3}$ as the
pressure is increased from ambient to 6 kbar.  Equivalently we can
assume that the density of moments remains equal to the Co density and
that the changes in the Curie constant reflect a reduction of $J$ from
1.3 to 0.9 with $P$. Our data could also be interpreted as a general
trend toward a stronger Kondo coupling with $P$ as has been shown to
be typical in the case of f-electron compounds\cite{maple}.

\subsection{The Specific Heat and Entropy\label{TheSpecificHeat}}
\begin{figure}[htb]
  \includegraphics[angle=90,width=3.0in,bb=60 300 560
  710,clip]{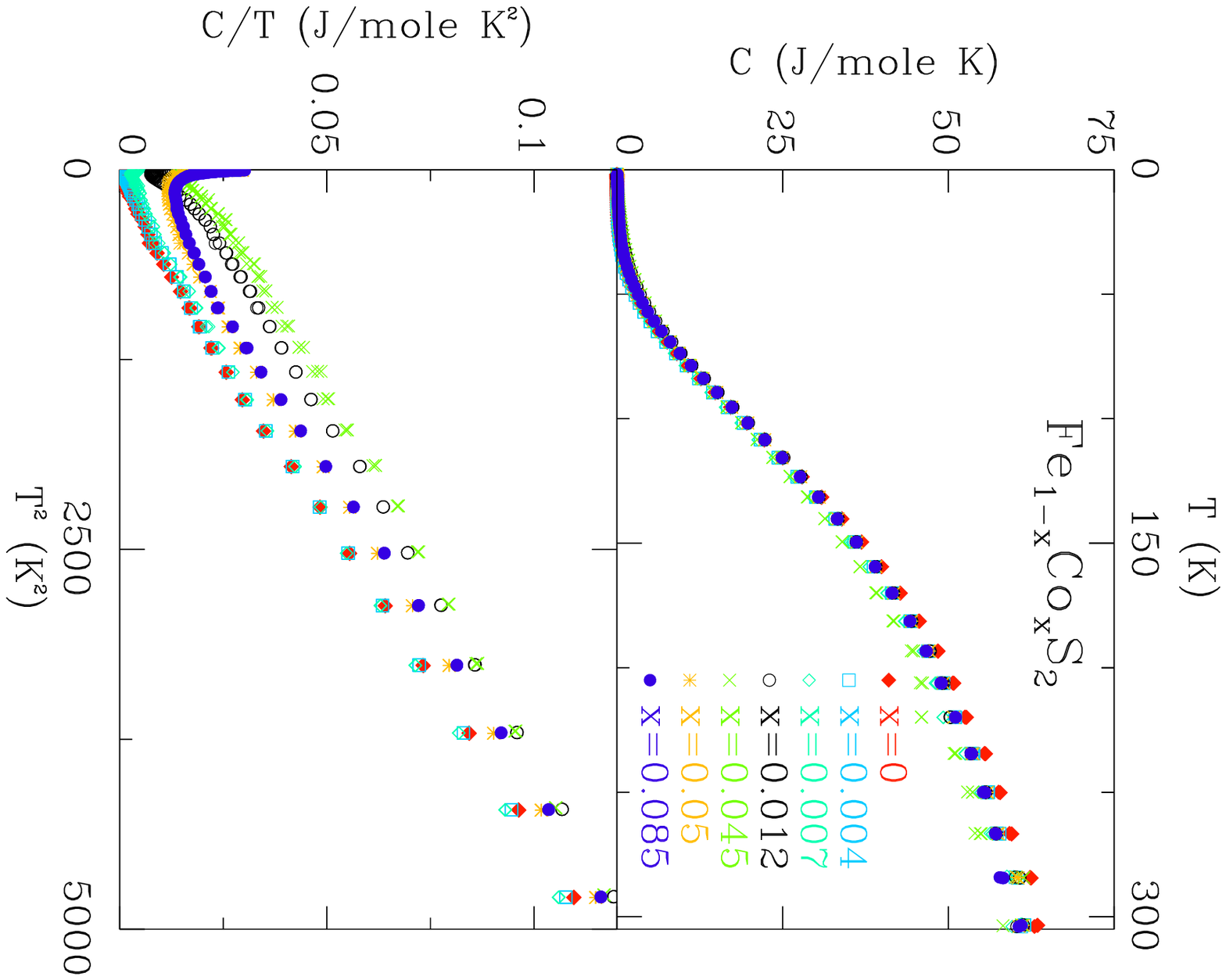}%
  \caption{\label{fig:sphpl} (Color online) Specific heat.  a) The
temperature, $T$, dependence of the specific heat, $C$ for seven
representative Fe$_{1-x}$Co$_x$S$_2$ crystals with $x$ identified in
the figure.  b) Specific heat divided by temperature, $C/T$, plotted
as a function of $T^2$ for temperatures between 2 and 70 K. Symbols
are the same as in frame a. }
\end{figure}

The specific heat, $C$ of our crystals was measured to characterize
the changes to the thermodynamic properties that occur upon Co doping
FeS$_2$. As can be seen in Fig.~\ref{fig:sphpl}, where $C$ is plotted
for 7 representative crystals, there is little change in $C(T)$ with
Co doping between 50 and 300 K. In frame b of the figure we
concentrate our attention on the low temperature specific heat where
changes in the electronic contributions are apparent. Here we plot the
data in the standard manner for characterizing the electronic
contributions, $C(T)/T$ vs.\ $T^2$, since in standard paramagnetic
metals the specific heat is expected to be accurately described by a
$C(T)/T = \gamma + \beta T^2$ form in the temperature range
displayed. In this equation $\gamma$ represents the coefficient of
electronic specific heat and is proportional to the density of
electronic states, while $\beta$ parameterizes the phonon contribution
to the specific heat at temperatures well below the Debye
temperature. While $\beta=2.17\times 10^{-5}$ J/mol K$^4$ describes
the phonon contribution of all our samples well, there are significant
changes to the electronic contribution to $C(T)$ with $x$. It is clear
that $\gamma$ of the nominally pure FeS$_2$ crystal is consistent with
zero and that $\gamma$ increases with $x$. However, there are
contributions to the specific heat that are not described by the
standard form. For example, at the lowest temperatures an upturn in
$C(T)/T$ with decreasing $T$ is apparent in all of the Co doped
samples. Several of the data sets also show an increased contribution
between 10 and 70 K that is not described by a simple constant. It is
most likely that the differences from the standard metallic form are
due to the formation of the magnetic ground state with Co doping that
was apparent in the magnetic susceptibility.

\begin{figure}[htb]
  \includegraphics[angle=90,width=3.0in,bb=60 300 560
  710,clip]{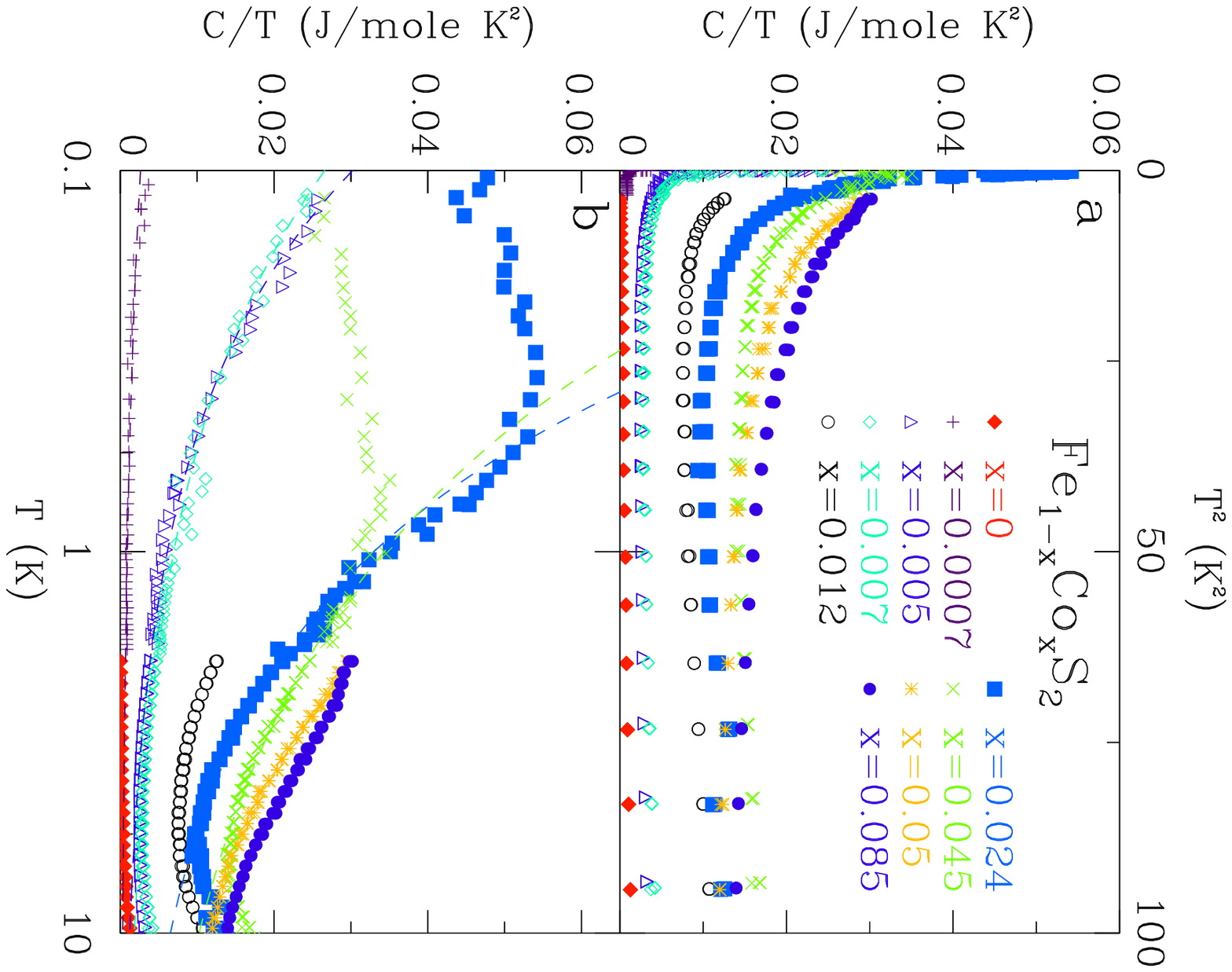}%
  \caption{\label{fig:sph2pl} (Color online) Low temperature specific
heat. a) The specific heat divided by temperature, $C/T$, plotted as a
function of $T^2$ below 10 K for nine representative crystals with
$x$ identified in the figure.  b) Specific heat divided by
temperature, $C/T$, plotted as a function of $T$ on a logarithmic
scale. Symbols are the same as in frame a. Dashed lines are fits of
the form $aT^{\alpha_C-1}$ to the data with an $\alpha_C$ of $-0.45 \pm
0.05$ for the $x=0.0007$, $-0.70 \pm 0.05$ for the $x=0.005$, $-0.60
\pm 0.05$ for the $x=0.007$, $-0.70 \pm 0.05$ for the $x=0.024$, and
$-0.50 \pm 0.05$ for the $x=0.045$ crystals. Note that the temperature
range for the fits was limited to $T>1 K$ for the $x=0.024$ and
$x=0.045$ crystals.}
\end{figure}

We expand the low-$T$ $C(T)/T$ data in Fig.~\ref{fig:sph2pl} to
highlight the changes that occur to the electronic contributions with
increasing $x$. In this figure the changes to $\gamma$ with $x$ above
1 K are made clearer along with the large increases in $C(T)/T$ that
occur below 1 K in the Co doped samples. The increases to $\gamma$
with $x$ are very large considering that Hall effect measurements
estimate the carrier concentrations are only 10 to 30\% of the Co
concentration in our samples\cite{guo}. Such a large density of
electronic states is commonly interpreted as a large electronic
effective mass. Our estimates for the carrier effective mass, $m^*$,
from the $C(T)/T$ data above 5 K range from 4 times the bare electron
mass, $m_e$, to over 40 $m_e$ for samples with $x$ just beyond
$x_c$. If we were to consider the large upturn in $C(T)/T$ below 1 K
to be electronic in origin, our estimates of $m^*$ increase to values
between 120 and 250 $m_e$ at 120 mK. Heavy mass carriers are rare in
transition metal compounds, however, several Fe-based semiconducting,
or semimetallic, materials have been reported to have substantial
carrier mass enhancement\cite{ditusa,nishino}. Crystals with $x<x_c$,
as determined from the AC magnetic susceptibility, display a $C(T)/T$
that increases continuously down to the lowest measurement
temperatures (0.1 K).  Those with $x>x_c$ display a maximum in
$C(T)/T$ at a temperature somewhat below $T_c$ as determined from the
temperature of the maximum in $\chi'$.

\begin{figure}[htb]
  \includegraphics[angle=90,width=3.0in,bb=60 380 490
  730,clip]{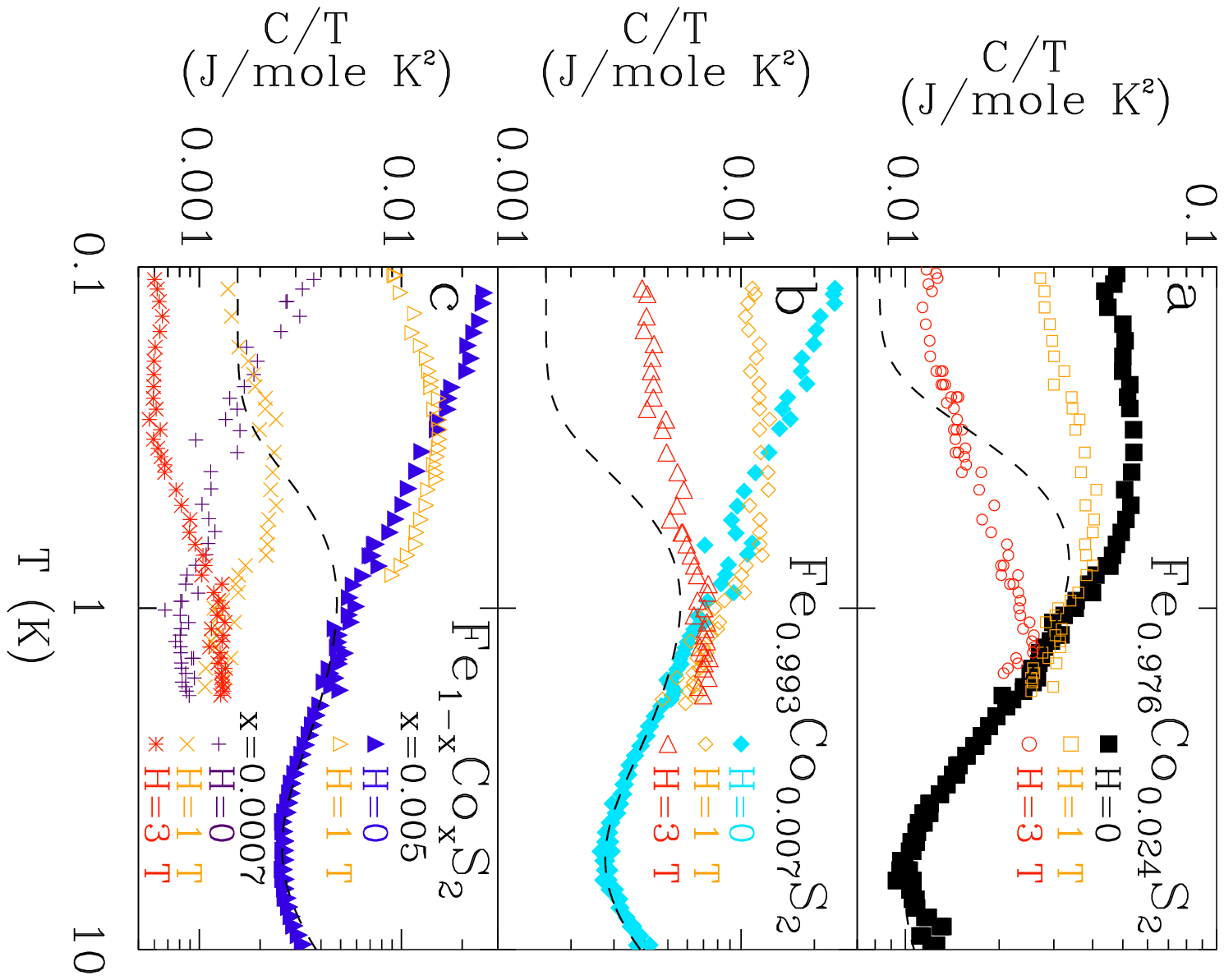}%
  \caption{\label{fig:sph4pl} (Color online) Magnetic field dependence
of the low temperature specific heat. a) The magnetic field, $H$, and
temperature dependence of the specific heat divided by temperature,
$C/T$ for an $x=0.024$ crystal at fields identified in the
figure. Dashed line is a fit of the Schottky
form{\protect{\cite{blundell}}} for a collection of free magnetic
moments in a magnetic field added to the phonon contribution to the
specific heat determined by $C$ of a nominally pure single crystal to
the $H=0$ $C/T$ data for $T>2$ K.  This fitting procedure was carried
out with the quantity $ng^2 J(J+1)$ held constant at the value
determined by fits of the Curie-Weiss form to the real part of the AC
magnetic susceptibility measured on the same crystal. The fitting
procedure resulted in best fit parameters of the effective field,
$H_{eff}=1.4$ T, a magnetic moment, $J=2.2$, and a density of magnetic
moments, $N = 9.5\times10^{19}$ cm$^{-3}$.  b) The $H$ and $T$ dependence
of $C/T$ of an $x=0.007$ single crystal at fields identified in the
figure. Dashed line is the a fit of the same form as in frame a to the
$H=0$ data. Best fit parameters were $H_{eff}=1.8$ T, $J=2.6$, and
$N=2.0\times10^{19}$ cm$^{-3}$. c) The $H$ and $T$ dependence of $C/T$ of
$x=0.005$ and $x=0.0007$ single crystals as identified in the figure
at fields identified in the figure. Dashed line is the a fit of the
same form as in frame a to the $x=0.005$, $H=0$ data. Best fit
parameters were $H_{eff}=1.88$ T, $J=2.4$, and $N=1.6\times10^{19}$
cm$^{-3}$.}
\end{figure}

In order to characterize the specific heat further, we have measured
the magnetic field dependence of $C(T)/T$ as shown in
Fig.~\ref{fig:sph4pl} for 4 of our single crystals. For all four of
the samples application of magnetic field decreases the low
temperature $C(T)/T$ while increasing slightly $C(T)/T$ above 1 K.
This behavior is qualitatively similar to the changes that occur upon
application of a magnetic field in simple doped semiconductors such as
Si:P\cite{lohneysen}. The low-$T$ specific heat of Si:P has been
modeled in terms of local moments induced by the doping that are
thought to form singlets such that the ground state of the system has
zero net moment\cite{BhattandLee,paalanen}. The effect of disorder is
to create a broad distribution of interactions between the local
moments. As a result, $C(T)/T$ increases as a small power-law at low
temperatures\cite{paalanen}. However, we do not believe such a
description to be relevant in the case of Fe$_{1-x}$Co$_x$S$_2$ since
our crystals have a nearly ferromagnetic, or ferromagnetically
ordered, ground state. Instead, we attribute the unusual temperature
dependence of $C(T)/T$ as being due to the interaction of a random
distribution of local moments together with the interaction of the
local moments and conduction electrons. As the magnetic field is
increased beyond the level of the interactions, $C(T)/T$ resembles a
sum of contributions from a Fermi gas of electrons (a constant
$\gamma$) and a separate contribution from effectively noninteracting
magnetic moments. The contribution from non-interacting moments in a
magnetic field is commonly referred to as a Schottky
anomaly\cite{blundell} and is demonstrated in the figure by the dashed
lines. We note that above 1 K the specific heat in zero field is well
described by this Schottky form added to a temperature independent
$\gamma$ and the phonon contribution taken as $C(T)/T$ of the
nominally pure FeS$_2$ crystal.

\clearpage

\begin{figure}[htb]
  \includegraphics[angle=90,width=3.0in,bb=95 300 350
  710,clip]{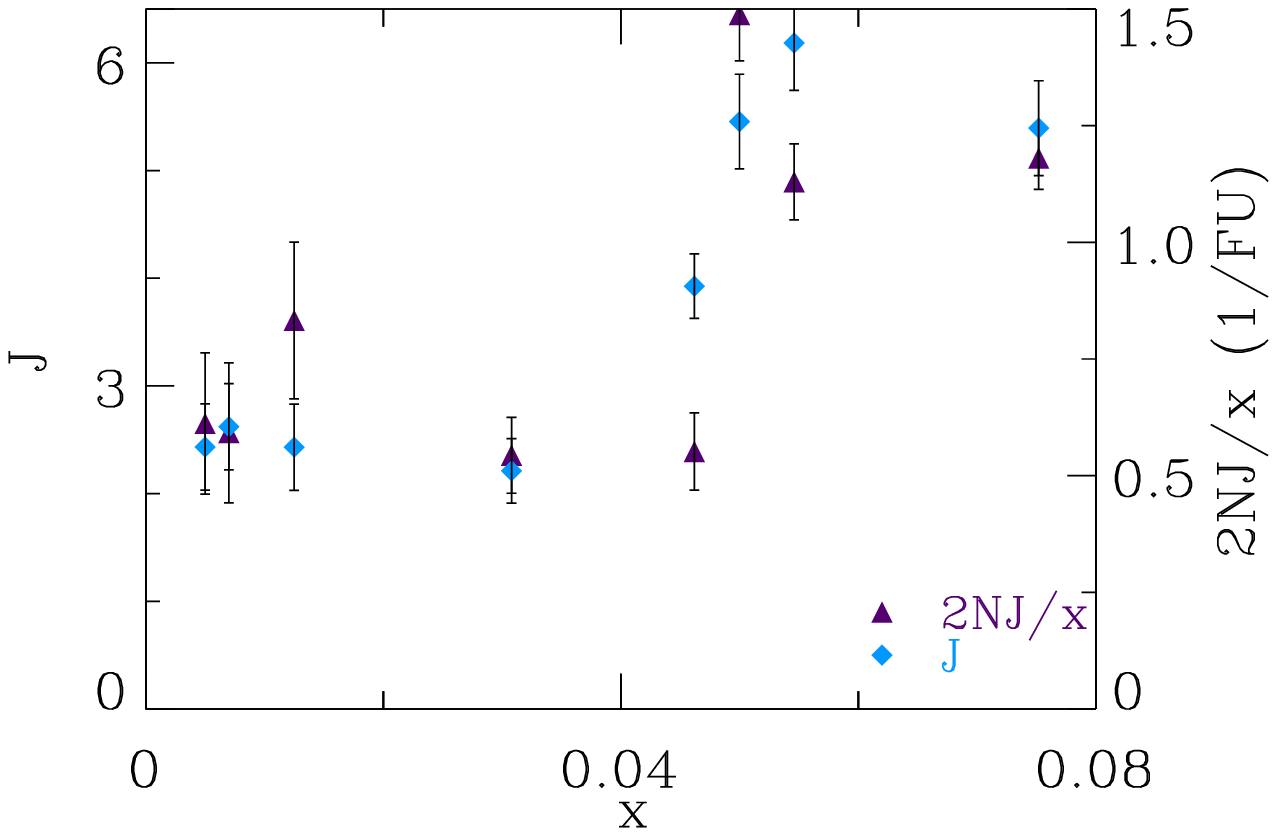}%
  \caption{\label{fig:mnj} (Color online) Magnetic moments from specific
heat and susceptibility analysis. Magnetic moment, $J$, and the
quantity $2NJ/x$ where $N$ is the density of magnetic moments of size
$J$ per Fe$_{1-x}$Co$_x$S$_2$ formula unit determined from fits to the
DC susceptibility and the specific heat above $T=2$ K
as described in text.  }
\end{figure}

We have checked that our conclusion of formation of ferromagnetic
cluster of magnetic moments at low-$T$ is consistent with our $C(T)$
measurements by quantitatively comparing $C(T)/T$ and $\chi_{DC}$. As
we stated above, the large Curie constants seen in Fig.~\ref{fig:tcpl}
are likely caused by the fluctuations of clusters of magnetic moments
and both the specific heat and the magnetic susceptibility are
sensitive to the density and size of the fluctuating
moments. Therefore, we have fit the specific heat at $T>2$ K as shown
in Fig.~\ref{fig:sph4pl} with the Schottky form\cite{blundell}
discussed above for a set of non-interacting magnetic moments
subjected to an effective molecular field. This contribution was added
to a $T$-independent term representing a Fermi liquid of conducting
electrons and the specific heat of the nominally pure FeS$_2$ crystal
as a model of the contribution to $C(T)$ from phonons. For this
fitting procedure we have held the the quantity $ng^2J(J+1)$ constant
at the value determined by $\chi_{DC}(T)$ and calculated the best fit
values of the effective molecular field, $H_{eff}$, $J$, and $n$.  In
Fig.~\ref{fig:mnj}b we present the results, plotting the best fit
values of $J$, which varied from less than 3 to 6 across our range of
$x$, and the parameter $2NJ/x$ where $N=n/n_{FU}$ and $n_{FU}$ is the
density of Fe$_{1-x}$Co$_x$S$_2$ formula units. We plot the quantity
$2NJ/x$ to check our assumption that each Co dopant donates a local
magnetic moment of $s=1/2$ to FeS$_2$. A value of $2NJ/x=1$
corresponds to agreement with this assumption and although there is
significant scatter in Fig.~\ref{fig:mnj}b, there is general
agreement. The best fit values of $H_{eff}$ varied between 1.4 and 2.6
T. We conclude that both $\chi'$ and $C(T)/T$ above 2 K are consistent
with fluctuating magnetic moments whose size is several times larger
than the magnetic moment of a single non-interacting Co dopant in an
FeS$_2$ lattice.

\begin{figure}[htb]
  \includegraphics[angle=90,width=3.0in,bb=280 350 560
  700,clip]{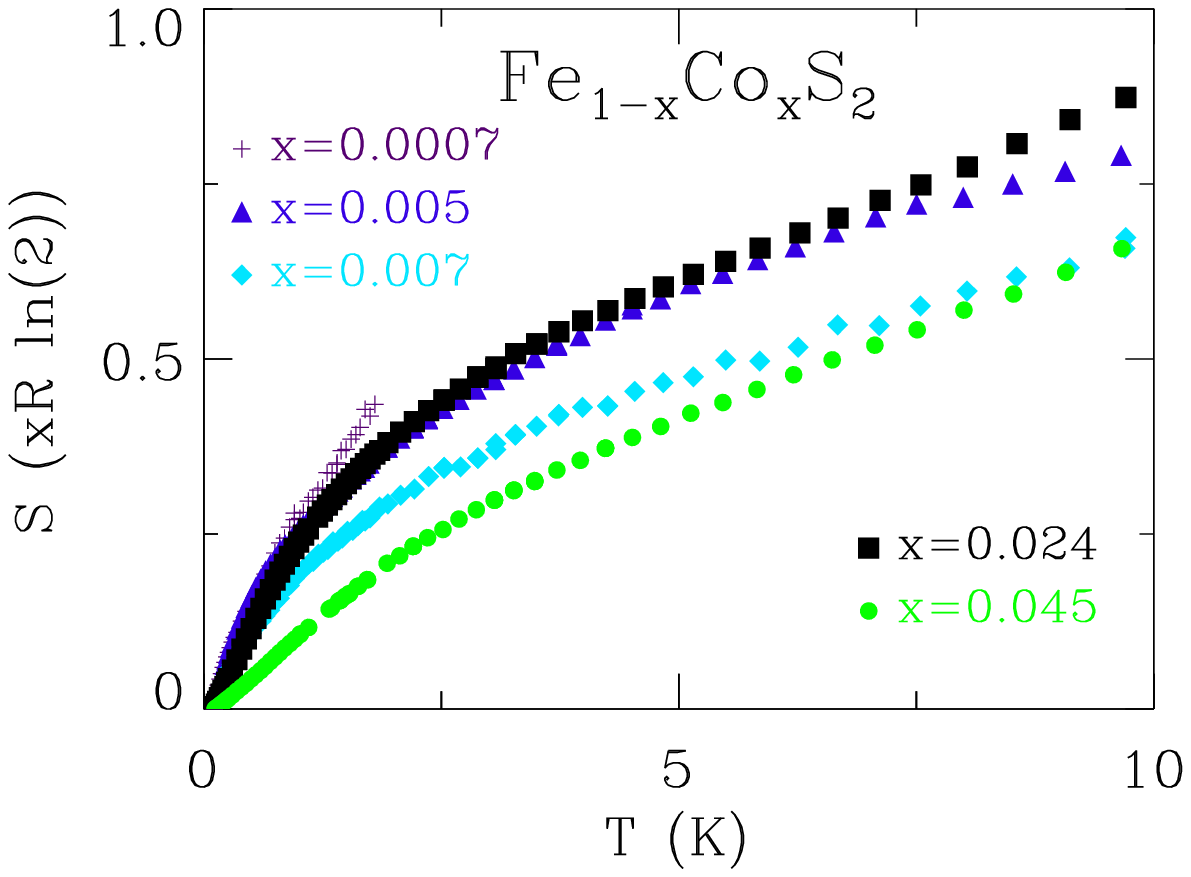}%
  \caption{\label{fig:entropy} (Color online) Entropy determined from
the specific heat. a) The temperature, $T$, dependence of the entropy,
$S$, determined by numerical integration of the $C/T$ data for five
representative single crystals with stoichiometry's identified in the
figure.  }
\end{figure}

The conclusion of magnetic cluster formation was further explored by
calculating the entropy, $S(T)$, from $C$ according to the relation
$\int_0^T C(T')/T' dT' = S(T)$ for the small number of crystals that
we have measured below 1 K. We have subtracted the phonon
contributions to $C(T)$ prior to performing the integrations and the
result of this procedure is shown in Fig.~\ref{fig:entropy}. Here $S$
approaches $xR\ln(2)$ only for temperatures well above 10 K. Thus,
there is significant entropy missing at 10 K consistent with the idea
that clusters of spins form at temperatures of this order.  Taking the
entropy at 7 K as a benchmark allows us to estimate the average
cluster size at this temperature. We make use of the form $S = (x/k)\:
R\: ln(2(k/2)+1)$ where $k$ is the average cluster size of spin 1/2
moments, presumably localized on Co sites, to make the estimate. We
find average cluster sizes that range from about 3 to 5 for the
samples where data below 1 K are available. If we assume an
exponentially decaying probability for clusters containing $N$
spin-1/2 Co impurities, we estimate that about 10\% of the cluster
sizes will contain 10 or more spins at this $T$.  The reduction of
entropy below $xR \ln{2}$ with cooling evident in
Fig.~\ref{fig:entropy} implies long ranged interaction between the
local magnetic moments induced by Co substitution. For these Co
concentration levels, and assuming that Co ions are substituted
randomly for Fe atoms in the FeS$_2$ crystal structure, we calculate
that the average distance between impurity sites to be at the level of
third nearest neighbors. Thus we estimate the range of interactions
between the magnetic moments to be at least 1 nm in order to produce
the reduction of entropy we observe below 10 K, as well as the
formation of magnetic clusters apparent in the magnetic susceptibility
to somewhat higher $T$.

\begin{figure}[htb]
  \includegraphics[angle=90,width=3.0in,bb=75 350 560
  700,clip]{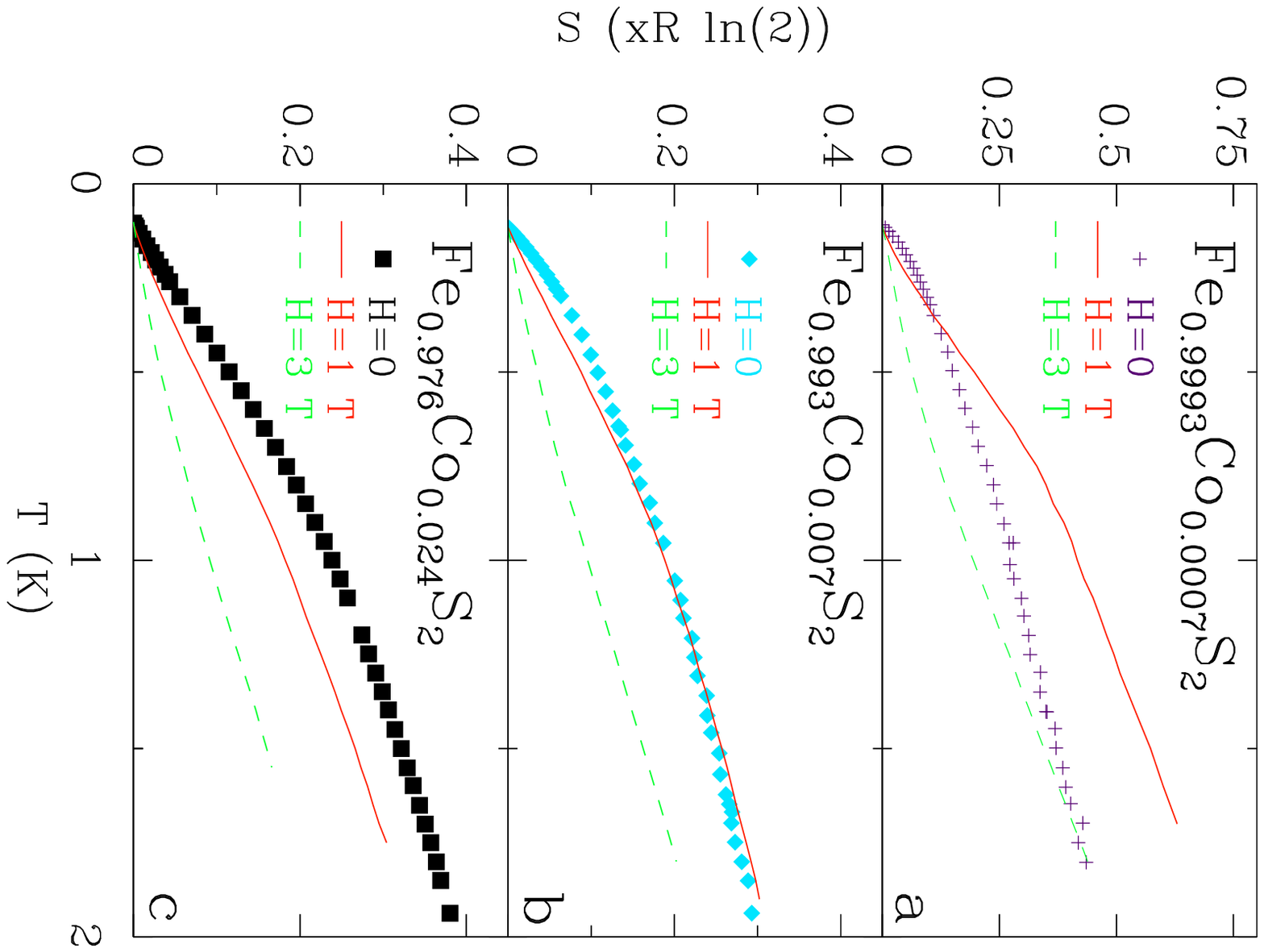}%
  \caption{\label{fig:entropy2} (Color online) Field dependence of the
entropy determined from the specific heat. The temperature, $T$,
dependence of the entropy, $S$, determined by numerical integration of
the $C/T$ data at magnetic fields, $H$ identified on the figure for a)
an $x=7\times10^{-4}$, b) an $x=0.007$, and c) an $x=0.024$ crystal.  }
\end{figure}

In addition, the changes to the entropy that occur with magnetic field
can be viewed in Fig.~\ref{fig:entropy2}. The sensitivity of $C(T)/T$
and $S$ to magnetic fields confirms the magnetic origin of the
specific heat in this temperature range. For very small $x$,
$x=7\times10^{-4}$ for example, increases in $S$ are observed below 2
K for $H=1$ T, a sign that there is still a significant contribution
to the entropy that we are not accessing because it lies below our
lowest measurement temperature. Thus, for small $x$, there is a
significant density of magnetic moments that interact with energy
scales of less than 100 mK. At larger $x$, $x\ge 0.007$, a suppression
of $S$ with field is observed from which we conclude that there are
very few magnetic moments with such small energies of
interaction. This is consistent with our conclusion that there is
significant cluster formation in these samples at temperatures above
those accessed in Fig.~\ref{fig:entropy2} and for our $x=0.024$
sample, a peak in $\chi'$ at 0.5 K that we associate with a disordered
ferromagnetic transition.

\section{Griffiths Behavior Near the Critical Concentration
\label{Griffneartc}}
As we noted in our previous discussion of both the AC and DC magnetic
susceptibility, the Curie constants determined from simple fits to the
temperature dependence were larger than is to be expected for
paramagnetically fluctuating spin-1/2 magnetic moments. That is, the
Curie constants displayed in Fig.~\ref{fig:tcpl} lie systematically
above the solid line representing the Curie constant expected when
each Co dopant donates a fluctuating $J=1/2$ magnetic moment. If we
focus our attention on those samples closest to $x_c$, we find Curie
constants 1.5 to 1.75 times larger than the simple estimate based on
the expectation that each Co dopant adds a single spin-1/2 moment and
the Curie law, $\chi(T) = CC/T$ with $CC=g^2\mu^2nJ(J+1)/3k_B$. This
is an indication that FM ordered clusters of magnetic moments are
forming below 100 K, where the Curie behavior becomes evident for this
dopant concentration. With the assumption that the magnetic moments
are coupled ferromagnetically, the size of the Curie constant would
indicate that on average each cluster contained between 2.5 and 3.2
spin-1/2 magnetic moments for $x\sim x_c$. In addition, we have
checked that this estimate is consistent with $C(T)$ above 2 K where
the data can be interpreted in terms of a Schottky anomaly associated
with entropy of fluctuating magnetic moments. This estimate of the
average size of fluctuating magnetic clusters is further supported by
estimates made from the entropy determined from the integral of
$C(T)/T$ (Fig.~\ref{fig:entropy}).

\begin{figure}[htb]
  \includegraphics[angle=90,width=3.2in,bb=60 260 540
  710,clip]{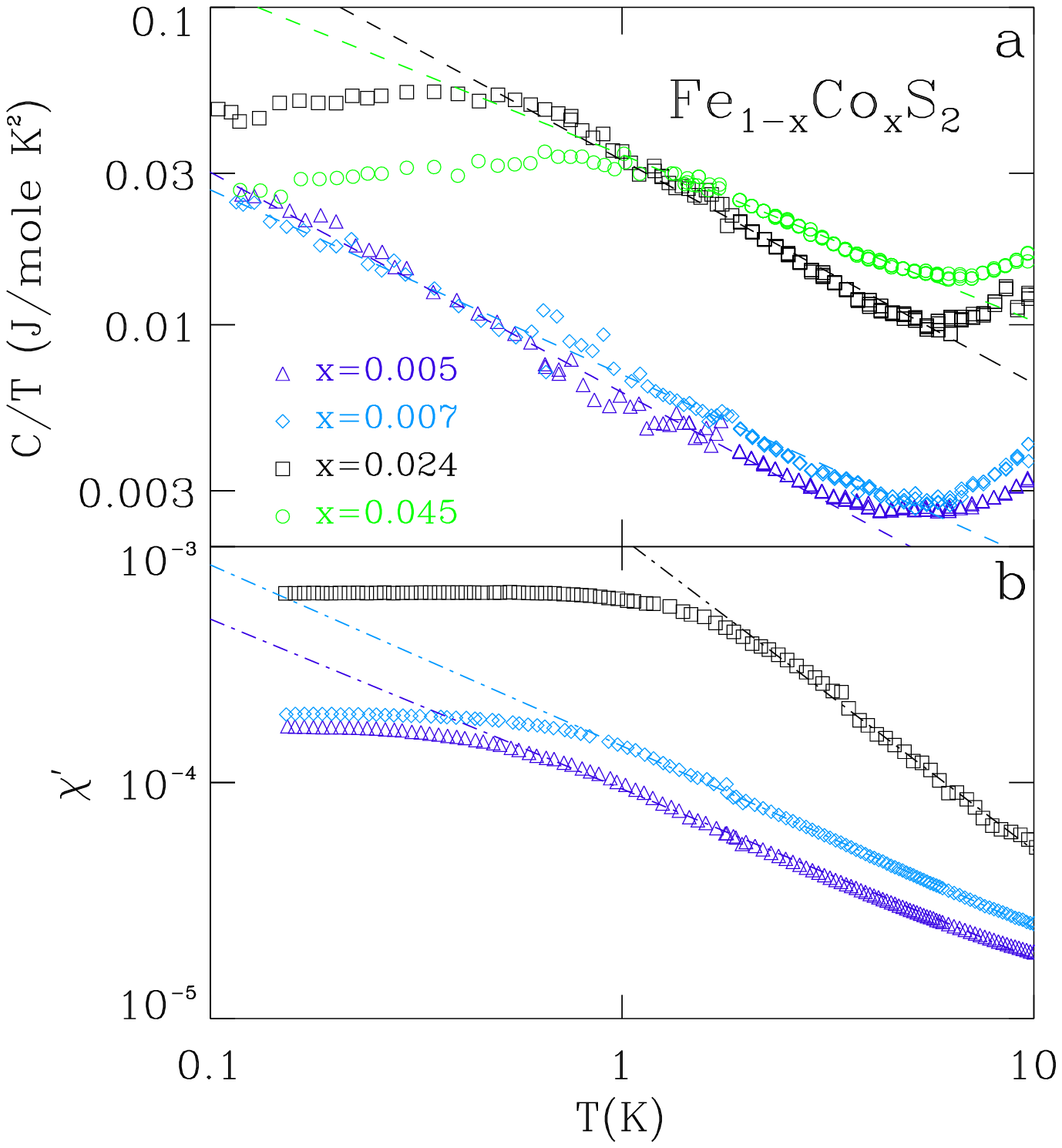}%
  \caption{\label{fig:sph3pl} (Color online) Low temperature specific
heat and magnetic susceptibility. a) The temperature dependence of the
specific heat divided by temperature, $C/T$ for several representative
crystals with $x$ identified in the figure. Dashed lines are fits to
$C/T$ to a power-law form, $C(T)/T \propto T^{\alpha_C -1}$ as in
Fig.~{\protect{\ref{fig:sph2pl}}}. b) real part of the AC magnetic
susceptibility, $\chi'$ for the same crystals as in frame a. Symbols
the same as in frame a. dashed dotted lines are fit of a power-law
form to $\chi'$, $\chi' \propto T^{\alpha_{\chi}-1}$ with exponents,
$\alpha_{\chi}$ of $0.25 \pm 0.1$ for the $x=0.005$, $0.22 \pm 0.1$
for the $x=0.007$ and $-0.3 \pm 0.2$ for the $x=0.024$ crystals. }
\end{figure}

Motivated by theoretical predictions about the importance of rare
regions in disordered magnetic systems\cite{vladrev,vojtarev}, we have
carefully examined the temperature and field dependent behavior of the
magnetization and specific heat of our samples with $x$ in proximity
to $x_c$ to search for such effects. We are not disappointed in that
unusual behavior is discovered. The most clear indication of the
importance of rare large clusters of magnetic moments comes from the
temperature dependence of the specific heat below 1 K shown in
Fig.~\ref{fig:sph2pl}b. Here $C(T)/T$ is seen to increase without any
suggestion of saturation down to 0.1 K for our samples closest to
$x_c$, $x=0.005$ and 0.007. As indicated by the quality of the fits of
a power-law form, represented by the dashed lines in the figure, as
well as the linearity of the data when plotted on a log-log scale in
Fig.~\ref{fig:sph3pl} a, these data are well represented by a $C(T)
\propto T^{\alpha_C-1}$ form with $\alpha_C$ between 0.3 and 0.45 for
over a decade in $T$. Similar power-law behavior is observed in
$\chi'$ over the same temperature range for the same crystals as can
be seen in Fig.~\ref{fig:sph3pl} b. Here $\chi'$ is well fit by a
$T^{\alpha_{\chi} - 1}$ form between 0.8 and 10 K with $\alpha_{\chi}$
between 0.2 and 0.25 for these samples. We note, however, that
$\chi'$, unlike $C(T)/T$ tends to saturate below 0.5 K.

\begin{figure}[htb]
  \includegraphics[angle=90,width=3.2in,bb=60 110 510
  710,clip]{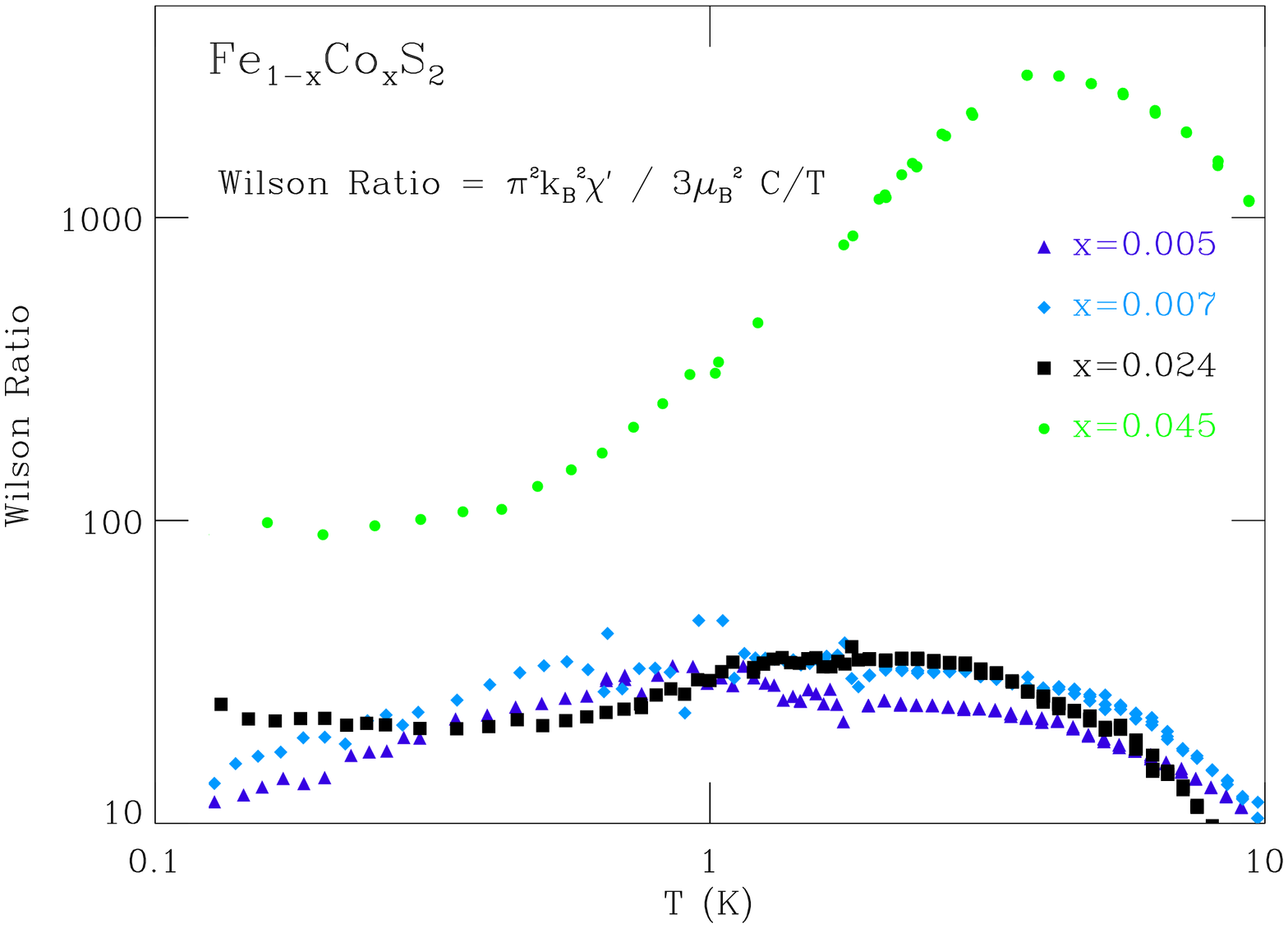}%
  \caption{\label{fig:wrplot} (Color online) Wilson Ratio. The
Wilson ratio, $\pi^2 k_B^2\chi' / 3\mu_B^2 C/T$, for crystals with $x$
identified in the figure.  }
\end{figure}

We make a direct comparison of the real part of the AC magnetic
susceptibility with $C(T)/T$ in Fig.~\ref{fig:wrplot} where the
Wilson ratio, defined as
\begin{equation}\pi^2\: k_B^2\: \chi' /\: 3 \mu_B^2\: C(T)/T
\label{wilsonratio} \end{equation}
is shown. In this figure we observe that the two samples with $x$
close to $x_c$ have a Wilson ratio somewhat larger than 10. This value
is similar to that observed near the critical point for magnetic
ordering in CePd$_{1-x}$Rh$_x$ which is thought to enter a novel
Kondo-cluster-glass phase and has properties compatible with the
quantum Griffiths phase scenario\cite{westerkamp}. Estimates for the
average sizes of the fluctuating magnetic moments from the Wilson
ratio of Fe$_{1-x}$Co$_x$S$_2$ yield about 3 to 4 spins consistent
with our other estimates\cite{vladrev}. We note that the Wilson ratio
at the lowest temperatures is not constant, but instead retains a
slight temperature dependence reflecting the observation that at
low-$T$ $C(T)/T$ continues to display a power-law temperature
dependence whereas $\chi'$ tends to saturate. We note that predictions
of the Wilson Ratio in Griffiths phases models indicate a weak,
logarithmic, temperature dependence\cite{vladrev}. For our samples
with $x>x_c$ the Wilson ratio becomes very large exceeding 1000 near
$T_c$ indicating a strong ferromagnetic tendency.

\begin{figure}[htb]
  \includegraphics[angle=90,width=3.2in,bb=60 260 540
  710,clip]{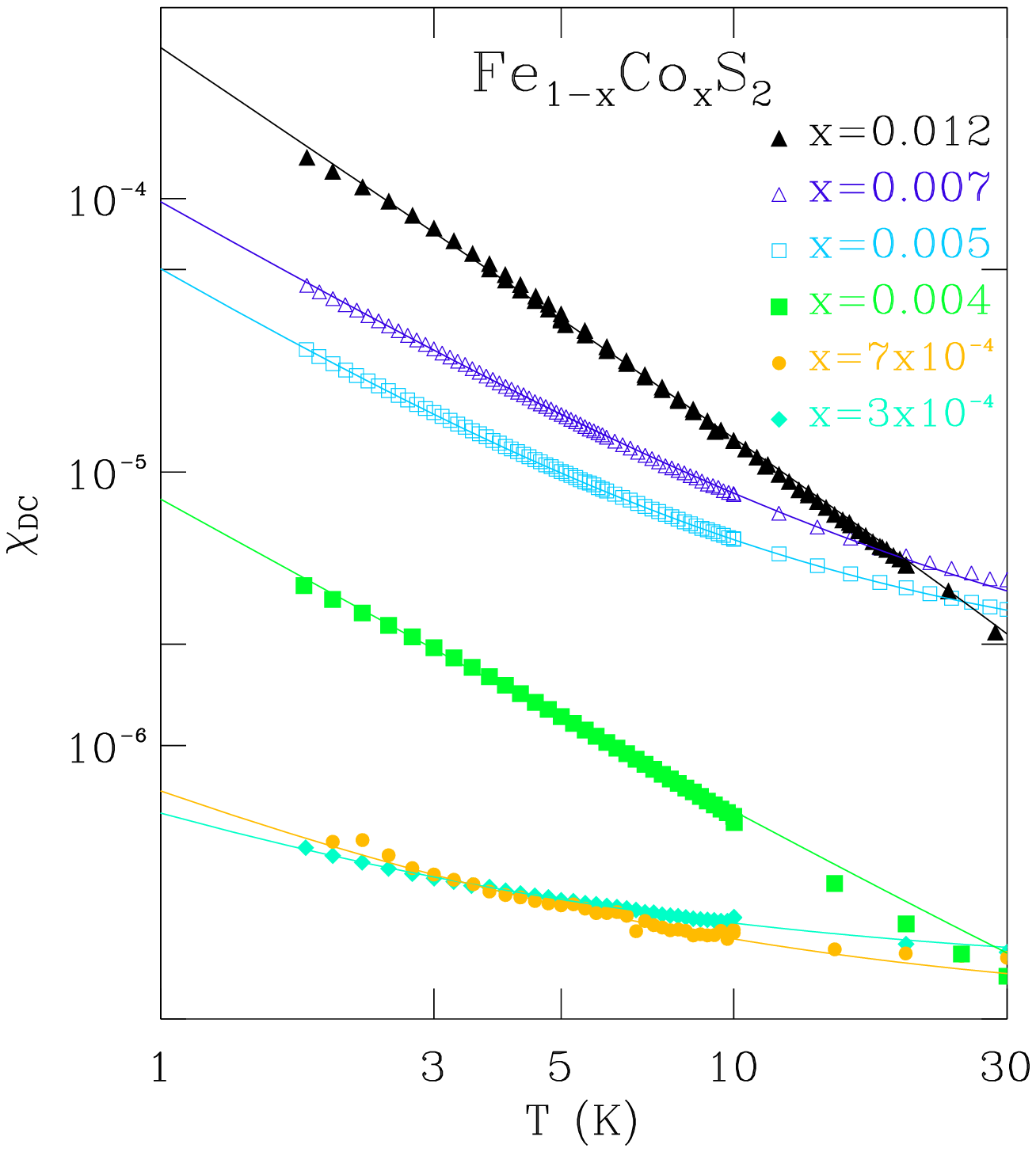}%
  \caption{\label{fig:chidcv3} (Color online) Power-law temperature
dependence of the DC magnetic susceptibility. The DC magnetic
susceptibility, $\chi_{DC}$, vs.\ temperature, $T$ for crystals with
$x \le x_c$ as identified in the figure. Data is the same as that shown in
Fig.~{\protect{\ref{fig:chidc}}}. Lines are fits to the form
$\chi_{DC} = \chi_{Pauli} + CC/T^{\alpha_{DC}-1}$ as described in the
text.}
\end{figure}

\begin{figure}[htb]
  \includegraphics[angle=90,width=3.2in,bb=60 260 505
  710,clip]{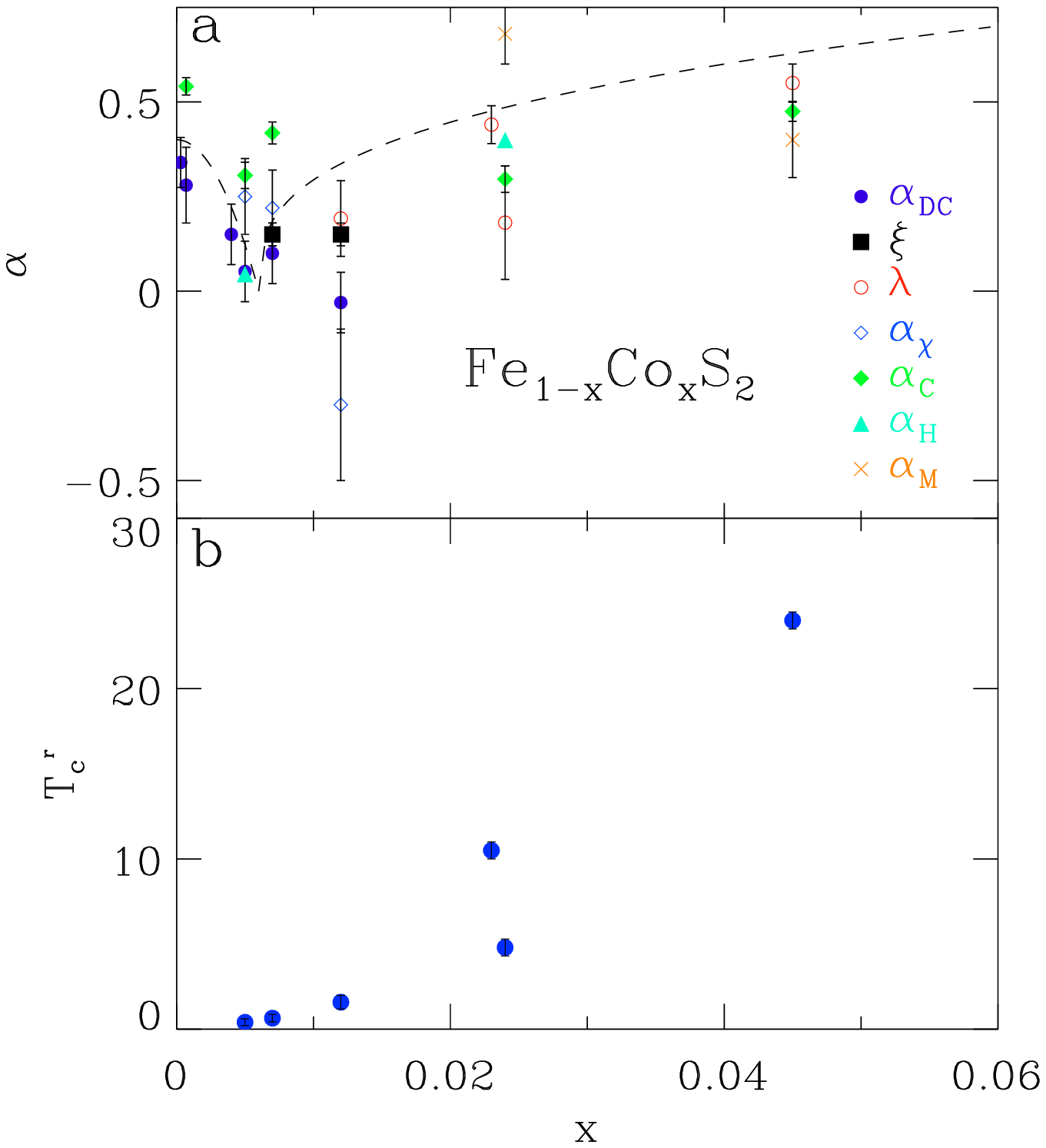}%
  \caption{\label{fig:chidcpwl} (Color online) Results of power-law
fits to the magnetic susceptibility and specific heat. (a) The
best-fit exponents from fits of simple power-law forms to the AC and
DC susceptibility and specific heat, $C$, temperature, $T$, and
magnetic field, $H$, dependence above the Curie temperature of
Fe$_{1-x}$Co$_x$S$_2$ for $3\times10^{-4} \le x \le 0.055$. Here
$\alpha_{DC}$ is the temperature exponent from fits of the form
$\chi_{DC} = cT^{\alpha_{DC} - 1}$ to the DC magnetic susceptibility
data below 10 K, $\xi$ determined from a scaling of $M(T,H)$ data
described in the text, $\lambda$ taken from fits of
Eq.~\ref{eq:grifchi} to $T$-dependence of the real-part of the AC
susceptibility, $\chi'$, $\alpha_\chi$ is from fits of $\chi'$ to the
form $\chi'=cT^{\alpha_{\chi}-1}$ between 0.8 and 10 K, $\alpha_C$
taken from fits of the form $C/T = cT^{\alpha_C-1}$ to the
$T$-dependence of the specific heat, $\alpha_H$ taken from fits of the
form $C/T = cH^{\alpha_H-1}$ to the $H$-dependence of $C/T$, and
$\alpha_M$ is taken from fits of the form $\chi'=cH^{\alpha-1}$ to the
magnetic field, $H$, dependence of $\chi'$. Dashed line is a
guide-to-the-eye. (b) Plot of the temperature characterizing the
maximum interaction between magnetic moments, $T_c^r$ taken from fits
of Eq.\ \ref{eq:grifchi} to the $T$-dependence of $\chi'$ for
$T>T_c$.}
\end{figure}

In addition, our finite-$H$ DC magnetic susceptibility measurements
indicate a small power-law temperature dependence below 10 K as can be
seen in Fig.~\ref{fig:chidcv3}. Although the temperature range is
restricted, for samples with $x<x_c$ a fit of the form $\chi_{DC}
\propto T^{\alpha_{DC}-1}$ results in small, between 0.4 and 0.05,
values of $\alpha_{DC}$. We note that these data were taken in
significant DC magnetic fields of between 1 kG and 1 T in order to
produce reasonable signal sizes for our SQUID magnetometery
measurements.  Even so, we find similar small power-law behavior with
the tendency for $\alpha_{DC}$ to fall toward zero near $x_c$. A
compilation of the power-laws determined from these different
measurements is presented in Fig.~\ref{fig:chidcpwl} where some
scatter in the values is apparent along with the tendency for larger
$\alpha$ values to be found in specific heat measurements than in
magnetic susceptibility measurements.

Our data and analysis outlined above have led us to the conclusion
that ferromagnetic clusters form at $T>T_c$. The sensitivity of our
susceptibility measurements to small increases in pressure is one of
the consequences of these weakly interacting ferromagnetically coupled
clusters. In addition, we observe small power-law dependencies of
$C(T)/T$, $\chi'$, and $\chi_{DC}$ at low-$T$ with changes in
temperature as was demonstrated in Fig.~\ref{fig:sph3pl} for $\chi'$,
in Figs.~\ref{fig:sph2pl} and \ref{fig:sph3pl} for the specific heat,
and is apparent in Fig.~\ref{fig:chidcv3} for $\chi_{DC}$. In each
case these quantities are observed to vary as $T^{\alpha -1}$ with $0
< \alpha < 0.6$. All of these power-laws have similar values although
there is significant scatter apparent. However, the similarities
between these power-laws, $\alpha_{M}$, $\alpha_{DC}$, and $\alpha_C$
for individual samples implies that a single physical mechanism may be
responsible.

\begin{figure}[htb]
  \includegraphics[angle=90,width=3.2in,bb=80 260 535
  710,clip]{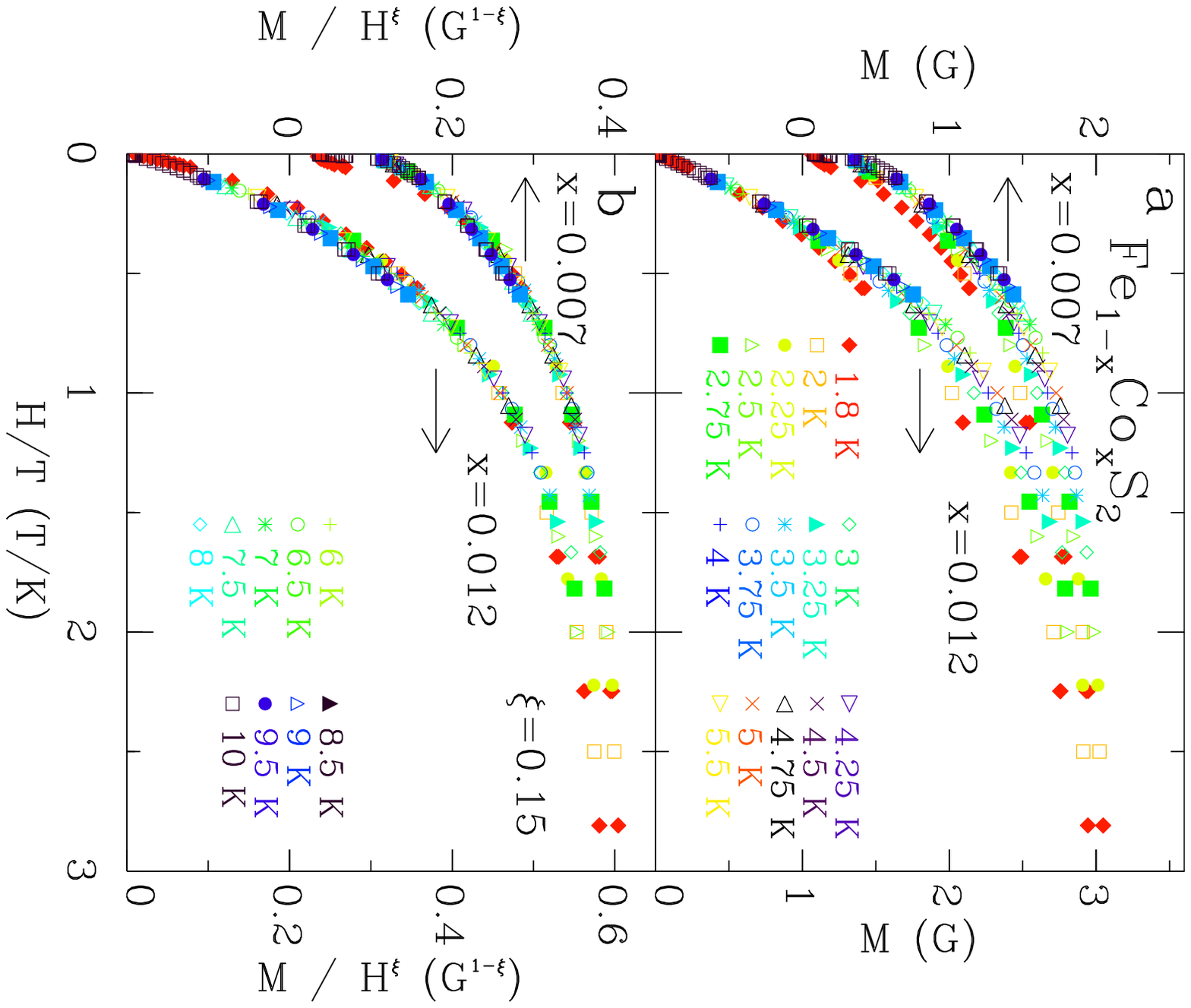}%
  \caption{\label{fig:vladsc} (Color online) Scaling plot of the
Magnetization. a) Plot of the magnetization, $M$, of
Fe$_{1-x}$Co$_x$S$_2$ crystals with $x=0.007$ (left side scale) and
$x=0.012$ (right side scale) as a function of magnetic field, $H$,
divided by temperature $T$, ($H/T$) to display the quality of the
standard paramagnetic scaling form $M(T,H)\propto F(H/T)$. Constant
temperature, $T$, sweeps are shown for $T$'s indicated in the frame
and in frame b for $1.8\le T \le 10$ K for $0 \le 5$ T. b) Plot of the
same $M(T,H)$ data as in frame a, divided by $H^{\xi}$ as a function
of $H/T$.  Reasonable scaling of the data is achieved for $\xi = 0.15
\pm 0.05$. }
\end{figure}

A clue as to the identity of the physical mechanism can be found in
the dependence of the magnetization on temperature and magnetic field,
$M(T,H)$. As we have noted above, the magnetic susceptibility does not
conform to Curie-Weiss dependence, instead being characterized by a
smaller power-law divergence at low-$T$.  This $T$-dependence suggests
a magnetization that is caused by independent fluctuators having a
power-law local energy distribution, $P(\Delta)\sim \Delta^{\xi-1}$,
where $\Delta$ is the local energy scale of the fluctuator.  The form
of the magnetization for freely fluctuating paramagnetic moments is
well known to follow the Curie law $M(T,H) = n g \mu_B J
B_J(g\mu_BH/k_BT)$, where $B_J$ is the Brillouin
function\cite{ashcroft} and thus scales as $M(T,H) \propto F(H/T)$. In
the case of the power-law energy distribution the magnetization has
been shown to scale as $M(T,H) = H^{\xi}F(H/T)$\cite{vladrev}.  As can
be seen in Fig.~\ref{fig:vladsc}a, which displays $M(T,H)$ between 1.8
and 10 K for our two crystals with $x$ closest to $x_c$, the the
standard scaling form does not scale our data satisfactorily. This is
to be expected for a system on the verge of magnetic ordering where
magnetic moments have significant interaction. However, by simply
allowing for a small value of $\xi$ we can scale all of our $H$ and
$T$ dependent data reasonably well as can be seen in frame b of the
figure ($\xi=0.15$). We note that $\xi$ is smaller than the values we
found for $\alpha_C$ and $\alpha_{DC}$, but still distinct from zero.
We note that this form for $M$ does not hold well for $x>0.04$ since
$M(H)$ increases sharply at low fields thus resembling more closely
$M(H)$ of ferromagnetic materials.

\begin{figure}[htb]
  \includegraphics[angle=90,width=3.3in,bb=85 100 540
  730,clip]{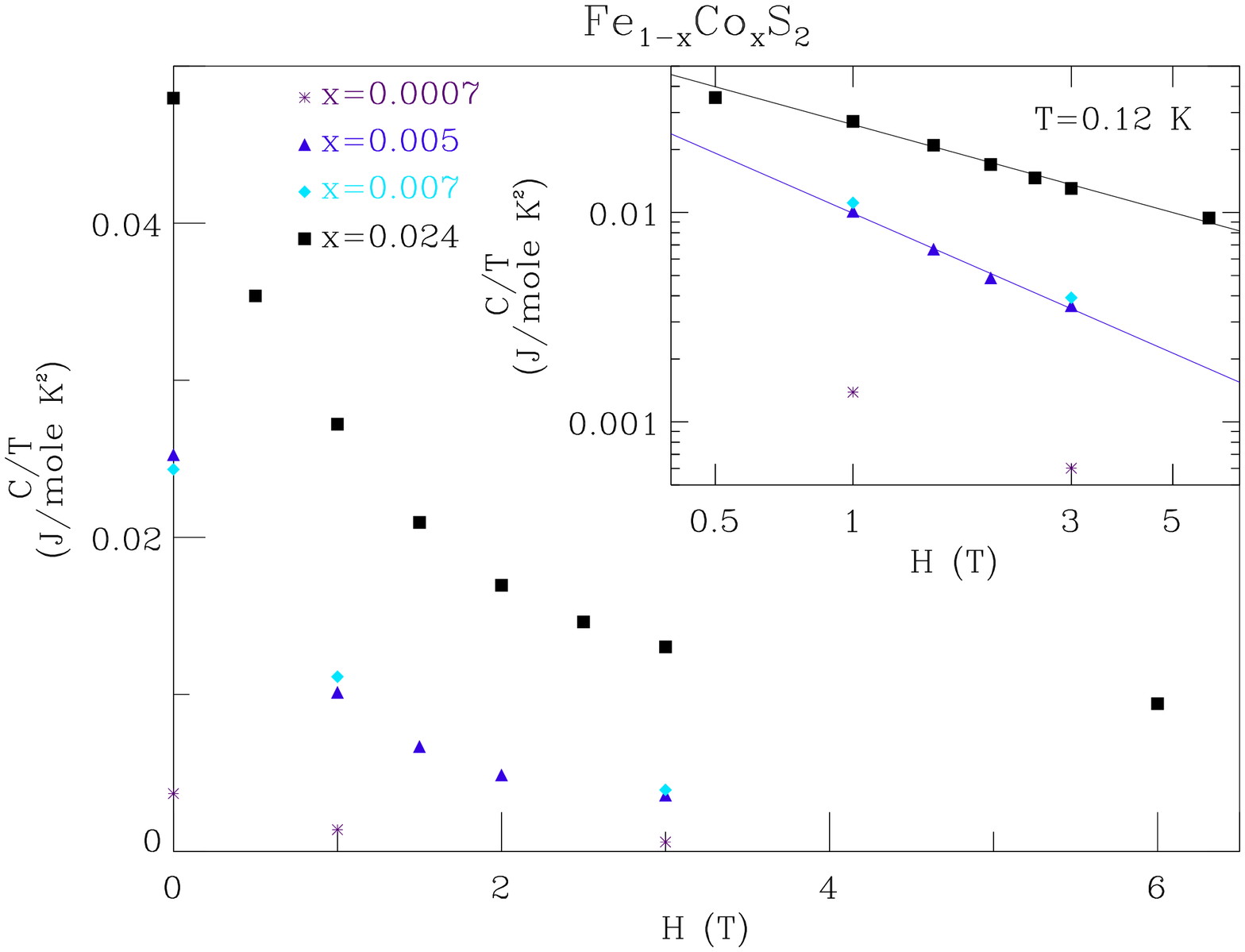}%
  \caption{\label{fig:sph5pl} (Color online) Magnetic field dependence
of the specific heat at $T=0.12$ K. The magnetic field, $H$,
dependence of the specific heat divided by temperature, $C/T$ for
several crystals with stoichiometry's identified in the figure.
Inset: The same $C(T,H)/T$ data plotted with logarithmic axes. The lines
are fits of a power-law form to $C/T$ data for the $x=0.005$ and
$x=0.024$ crystals. Best fit values of the exponents are $-1$ for
the $x=0.005$ crystal and $-0.6$ for the $x=0.024$ crystal.}
\end{figure}

In addition to this unusual field and temperature dependence of $M$,
our specific heat measurements also reveal a strong sensitivity to
magnetic fields as we demonstrated in Fig.~\ref{fig:sph4pl}. Although
we found that above 2 K the $T$ and $H$ dependence of $C/T$ appears
consistent with a Schottky anomaly form, $C/T$ below 1 K is not, even
when a 3 T magnetic field is applied. This is emphasized by the
magnetic field dependence of the low-$T$ specific heat which is
explicitly displayed for $T=120$ mK Fig.~\ref{fig:sph5pl}. Here
moderately sized fields are seen to suppress $C(T)/T$
dramatically. The power-law like decrease in $C(T)/T$ is displayed in
the inset where the lines are power-laws in field, $C(T)/T \propto
H^{\alpha_H-1}$ with $\alpha_H = 0$ for the $x=0.005$ crystal and
$0.4$ for the $x=0.024$ crystal. We note that for a Schottky anomaly
$C(T)/T$ decreases exponentially in $H$ for fields where
$g\mu_BH>k_BT$ which is inconsistent with our data\cite{blundell}.

Although there are many physical realizations of this scaling form
for the magnetization we observe (independent fluctuators with a
power-law energy distribution), and small power-law $T$ dependent
magnetic susceptibilities have been known to describe doped
semiconductors very near insulator-to-metal transitions, our
identification of magnetic clusters forming in this doped
semiconducting system presents an obvious candidate explanation for
all of the data presented thus far. We posit that our data lead
naturally to the conclusion that Griffiths phase formation is the most
likely physical mechanism responsible for the unusual properties we
measure for samples in proximity to the critical concentration for
magnetic ordering. 

Magnetic Griffiths phenomena have been used to describe disordered
magnetic systems theoretically in the case where the disorder is
sufficient to cause inhomogeneity in the formation of the magnetic
ground state\cite{Griffiths}. In the vicinity of the quantum critical
point, droplets of the magnetically ordered phase are thought to form
due to the inhomogeneities inherent to a chemically substituted
material. These droplets reside within the paramagnetic phase and can
be exponentially rare if the statistics are Poissonian so that $P(V)
\sim e^{-cV}$, where P is the probability of having a droplet of
volume $V$ and $c$ is related to the strength of the
disorder\cite{vladrev}. Since Fe$_{1-x}$Co$_x$S$_2$ has a
ferromagnetic ground state for $x>x_c$, the droplets most likely carry
a net magnetic moment simply proportional to the size of the
droplet. These magnetic moments will have a finite probability of
tunneling to nearby excited magnetization states. The tunneling rate
for such a droplet is predicted to be exponential in the number of
spins that form the droplet, such that $\Delta \sim \omega_0 e^{-bV}$
where b is a constant related to the microscopic tunneling mechanism,
and $\omega_0$ is a frequency cutoff\cite{vladrev}.  Since the
tunneling rate is proportional to the energy splitting of the
different magnetization states the distribution of energy splittings
can be written as $P(\Delta) \sim \int dV P(V) \delta [\Delta -
\omega_0 e^{-bV}] \sim \Delta^{\xi -1}$, where $\xi = c/b$, giving us
our power-law distribution of energy levels suggested by the scaling
of our $M(T,H)$ data. With this form for the energy splittings,
power-law forms for the magnetic susceptibility and specific heat
follow by considering all clusters with $\Delta > k_BT$ to be frozen
such that they do not contribute to the susceptibility or
entropy\cite{vladrev}. Thus, a modified Curie form $\chi'(T) \propto
n(T)/T$, where $n(T)$ is the density of clusters with $\Delta < k_BT$
is predicted\cite{vladrev,vojtarev}.  $n(T)$ can be easily calculated
from $P(\Delta)$ as $n(T) \propto \int_{0}^{T}P(\Delta) d\Delta
\propto T^{\alpha}$ and thus, $\chi'(T) \propto T^{\alpha-1}$. In
addition, $S(T) \propto n(T) R \ln{(2J+1)} \propto T^{\alpha}$ and
since $C(T)/T = dS/dT$ we have $C(T)/T \propto T^{1-\alpha}$.
Therefore, the Griffiths phase phenomenology leads to a form of the
magnetization, susceptibility, and, $C(T)/T$ similar to that found in
our experiments.

There has been some discussion in the literature about the ability of
magnetic clusters to tunnel into different magnetization states in the
presence of itinerant electrons. The presence of the conduction
electrons which couple to the local magnetic moments via the Kondo
effect cause dissipative effects on the tunneling of magnetic
clusters. The dissipation limits the size of clusters allowed to
tunnel into opposing magnetization states thus cutting off the
singular behavior in the models\cite{millis2}.  Millis et al.\ argue
that in heavy fermion metals the the electrons are sufficiently well
coupled to local magnetic moments, with Kondo temperatures of order
100 K, so that Griffiths phases are unlikely to be observed in these
systems\cite{millis2}. The idea being that the dissipation is relevant
at the scale of the Kondo temperature, thus the temperature region
where Griffiths phases may be observed is very small. However, Castro
Neto and Jones argue that in the region of magnetic clusters the
coupling of conduction electrons to the magnetic moments is not
important since in these regions the electron density of states is not
renormalized by the Kondo effect\cite{CastroNeto,castro2}.

In Fe$_{1-x}$Co$_x$S$_2$ the relevant energy scales are such that
Griffiths phases may be observed over a much wider temperature
scale. Here, the Kondo coupling scale is only of order 2 K in the
region of the critical Co concentration for ferromagnetism. The Kondo
scale is likely so small because the density of electron states, in
the approximation of a spherical electron pocket, is proportional to
$n_e^{2/3}$ where $n_e$ is the electron density which is more than 2
orders of magnitude smaller, for the $x$ range investigated here, than
in most heavy fermion metals\cite{guo}. In addition, the disorder
inherent in a doped semiconducting system results in a high scattering
rate of the electrons further reducing their ability to screen local
moments. Therefore, we expect the dissipation due to a conducting gas
of electrons to be a much smaller effect than in the heavy fermion
metals where electron densities are much larger and the strong Kondo
coupling renormalizes the electron DOS to very high levels. In
addition, the relevant energy scale for local moment interactions is
commonly estimated by the Curie temperature of the clean magnetic
system without dilution. In this case the clean ferromagnet is CoS$_2$
and this sets the largest energy scale for local moment interactions
at 120 K. Estimates for the average interaction energy of local
moments include the Weiss temperature which we have measured via the
magnetic susceptibility and is displayed in Fig.~\ref{fig:tcpl}. From
the average and maximum local moment interaction scales compared with
the Kondo temperature scale in our materials, we conclude that
dissipative effects of the conduction electron gas should be much
smaller in Fe$_{1-x}$Co$_x$S$_2$ than in heavy fermion
metals. However, rounding of the transition due to Kondo coupling of
the local moments and the conducting electrons may be relevant at the
lowest temperatures.

\section{Finite Temperature Magnetic Transitions; $x>x_c$
\label{finitetc}}
The figures and discussion above give a general description of
Fe$_{1-x}$Co$_x$S$_2$ as it evolves from a diamagnetic insulator into
a ferromagnetic metal. These data establish a magnetic transition that
can be observed via the AC magnetic susceptibility down to very low
temperatures at very small $x$ . It is clear from our data that
magnetic clusters form above $T_c$ and an inhomogeneous magnetic state
below $T_c$. In this section we explore in more detail the behavior of
our Fe$_{1-x}$Co$_x$S$_2$ crystals having a magnetic phase transition,
or perhaps a magnetic glass freezing, at finite temperatures. We
examine the frequency and magnetic field dependence of the magnetic
susceptibility both above and below the ordering to compare with
typical spin glasses and disordered FMs to gain insight on how such a
low density disordered metal establishes a magnetic ground state.

\begin{figure}[htb]
  \includegraphics[angle=90,width=3.2in,bb=70 245 500
  660,clip]{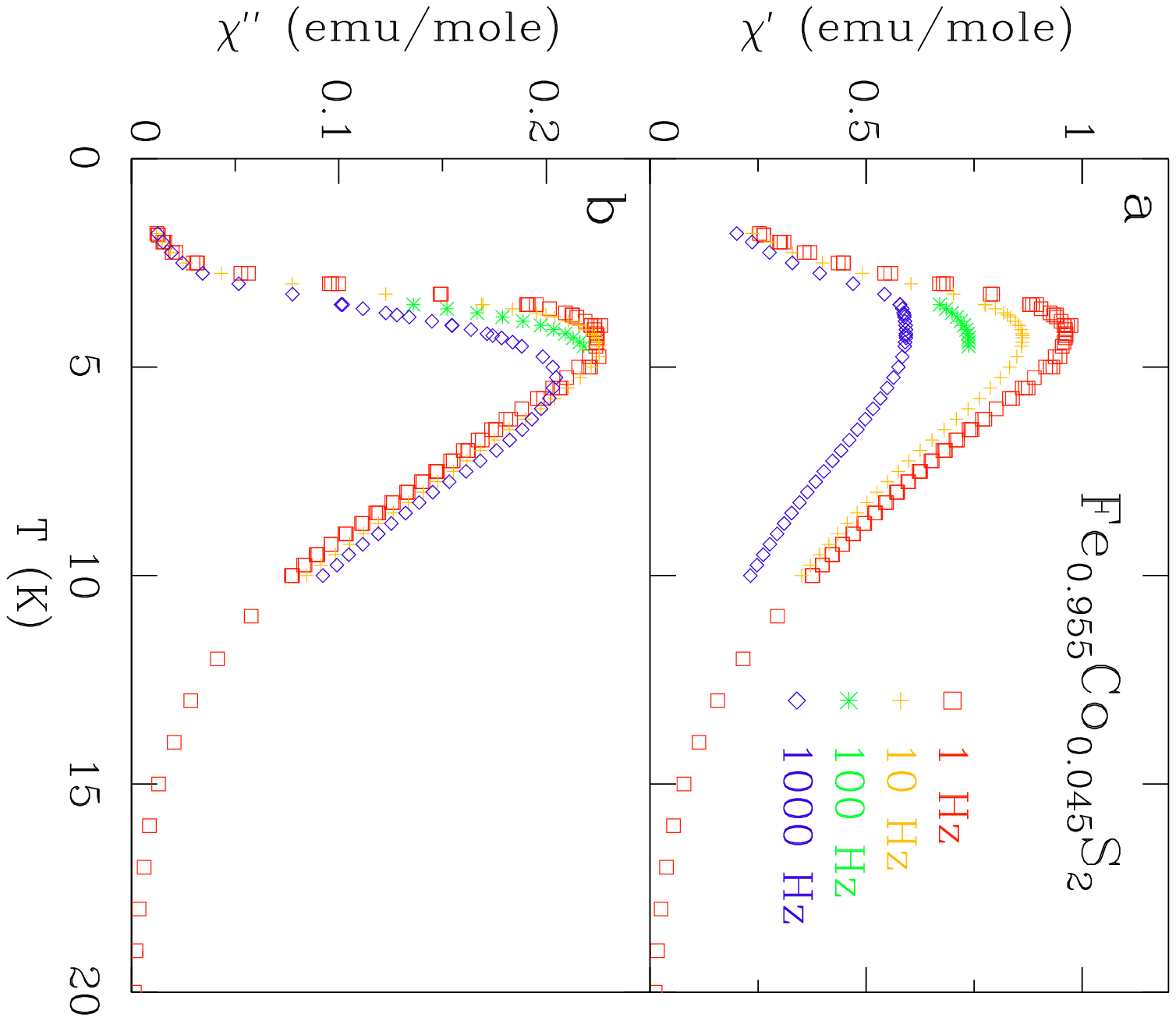}%
  \caption{\label{fig:chifreqtemp} (Color online) Frequency dependence of
the AC susceptibility. The frequency and temperature dependence of the
(a) real part, $\chi'$, and (b) imaginary part, $\chi''$, of the AC
magnetic susceptibility at zero DC field for several $x=0.045$ single
crystals. Measurement frequencies were varied between 1 and 1000 Hz as
identified in the figure. The excitation field was 1 Oe. }
\end{figure}

 To this end we have carefully measured the $f$-dependence of $\chi'$
and $\chi''$ of a few of our samples with $x>x_c$ to probe the low
frequency dynamics of the highly paramagnetic state above $T_c$ as
well as the magnetic state below $T_c$. In Fig.~\ref{fig:chifreqtemp}
$\chi'$ and $\chi''$ of our $x=0.045$ sample is displayed for
frequencies between 1 and 1000 Hz. Here, we observe significant
decreases in the magnitude of both $\chi'$ and $\chi''$, very small
changes to the temperature of the maximum in $\chi'$ (at most 0.1 K
increase), and a larger increase of the temperature of the maximum in
$\chi''$ with measurement frequency. We observe no indications of the
onset of significant skin effects and eddy current absorption which
cause a steady increase in the out-of-phase susceptibility ($\chi''$)
with $f$ despite the large out-of-phase component of the signal.

The frequency dependence we measure is similar to the frequency
dependent changes that occur in the archetypal spin glass
CuMn\cite{mulder} with a relative shift of $T_c$ per decade of
measurement frequency of less than 1\%. However there are with two
important differences with typical spin glass behavior that draw our
attention. First, we observe much larger changes in the magnitudes of
both $\chi'$ and $\chi''$ with frequency than is typical for a
metallic spin glass. Second, the temperature range over which we
observe a suppression of $\chi'$ is from at least 0.5 $T_c$ to $\sim
2T_c$, whereas for CuMn of similar dopant concentrations there is no
measurable $f$-dependence of $\chi'$ for reduced temperatures, $(T -
T_f)/T_f > 0.05$, where $T_f$ is the spin glass freezing
temperature\cite{mulder}.

\begin{figure}[htb]
  \includegraphics[angle=90,width=3.2in,bb=70 290 500
  665,clip]{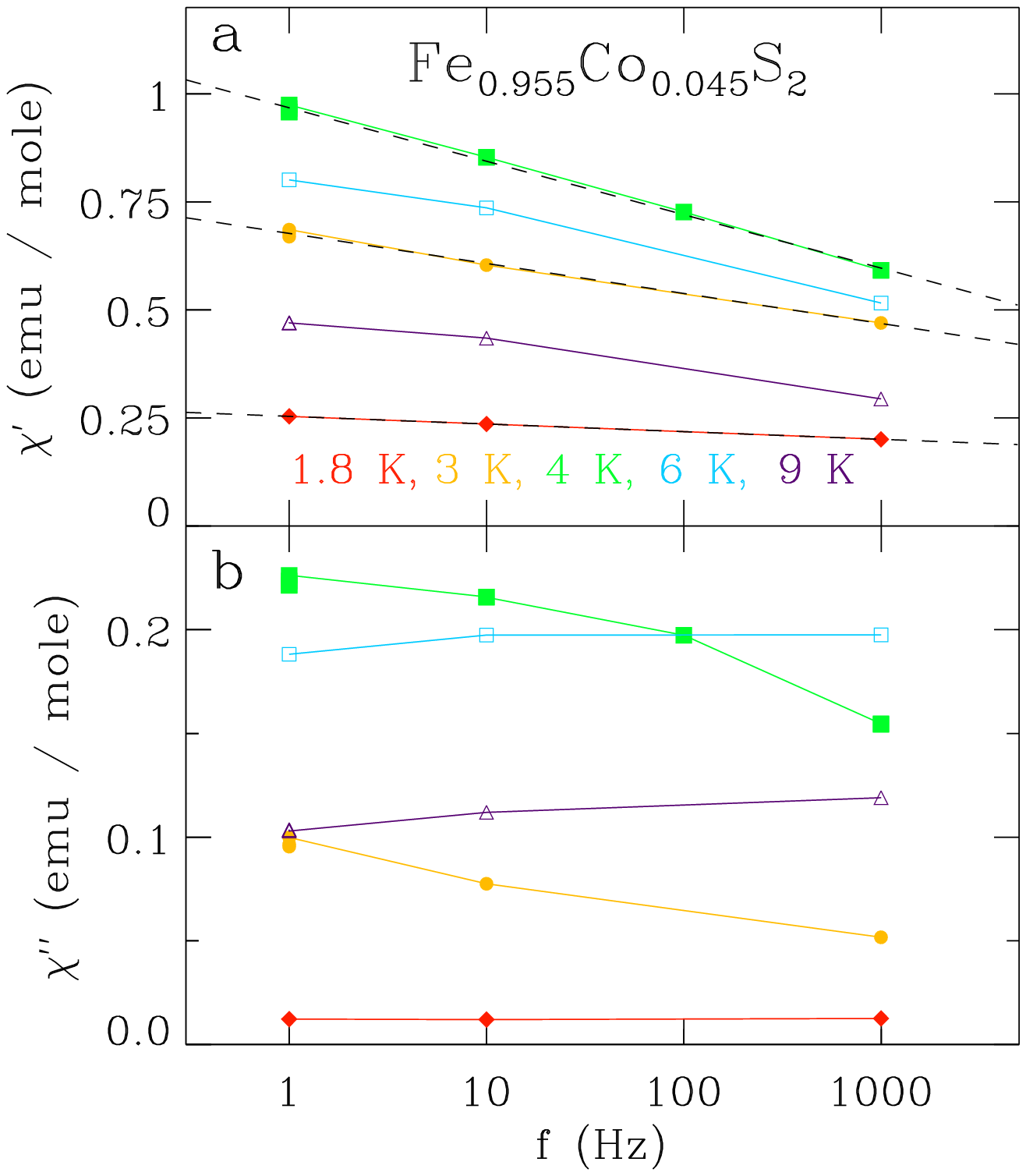}%
  \caption{\label{fig:chitempfreq} (Color online) Frequency dependence
of the AC susceptibility. The frequency dependence of the real part,
$\chi'$, (a) and imaginary part, $\chi''$, (b) of the AC magnetic
susceptibility at zero DC field for several $x=0.045$ single
crystals. Temperatures shown are 1.8 K (red diamonds), 3 K (orange
bullets), 4 K (green filled squares), 6 K (blue boxes), and 9 K
(violet triangles). The excitation field was 1 Oe. Dashed black lines
in frame a are fits to a logarithmic frequency dependence of $\chi'$.}
\end{figure}

The frequency dependence of $\chi'$ and $\chi''$ for several
temperatures is explicitly demonstrated in Fig.~\ref{fig:chitempfreq}
for temperatures near $T_c$. While $\chi'$ decreases monotonically
with frequency $\chi''$ is characterized by a broad peak which moves
through the frequency window of our measurement from low $f$ to high
$f$ with temperature. At $T<T_c$, $\chi'(f)$ is well described by a
logarithmic dependence, as demonstrated in the figure by the dashed
lines, with an extrapolated zero crossing between $10^8$ and $10^{14}$
Hz. This dependence has also been observed in insulating spin glasses
such as Ising-like LiHo$_x$Y$_{1-x}$F$_4$\cite{ancona}. In addition,
$\chi''$ has very little $f$-dependence in the same temperature range
so that the dynamic response of the system below $T_c$ is much like
that of a spin glass where fluctuations occur on all long time
scales\cite{binder}. Thus, the magnetic dynamics at these low
frequencies resemble that of common spin glass systems, with the added
feature that the time dependence extends to temperatures far above
$T_c$.

\begin{figure}[htb]
  \includegraphics[angle=90,width=3.2in,bb=70 315 500
  710,clip]{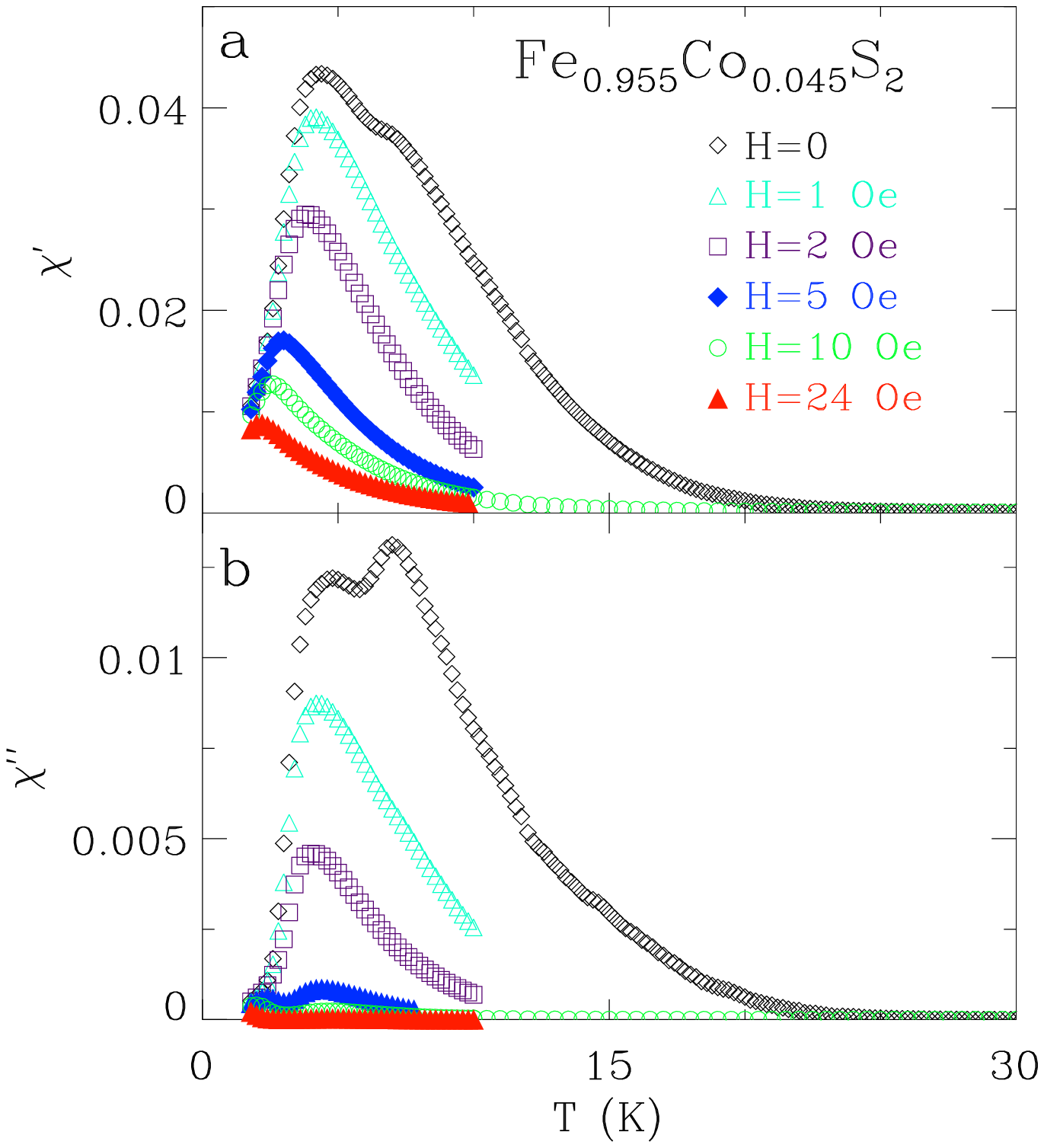}%
  \caption{\label{fig:chifield3pl} (Color online) Dramatic field
dependence of the AC magnetic susceptibility. The magnetic field, $H$,
and temperature, $T$ dependence of (a) the real part, $\chi'$, and (b)
imaginary part, $\chi''$, of the AC susceptibility, for the same
sample as in Fig.~ \protect{\ref{fig:chifreqtemp}} at fields identified
in the figure. Data taken with an excitation field of 1 Oe at a
frequency of 1 Hz.}
\end{figure}

In addition to the enhanced sensitivity to measurement frequency near
$T_c$ our crystals display an enhanced sensitivity to external
magnetic fields in the same temperature range.  This is demonstrated
in Fig.~\ref{fig:chifield3pl} for an $x=0.045$ sample consisting of
several small crystals. We note that the data displayed in
Fig.~\ref{fig:chifield3pl}a was previously published\cite{guo} and is
reproduced here for completeness and comparison of $\chi'$ and
$\chi''$. In this experiment we have carefully minimized the DC
magnetic field for our $H=0$ scan which reveals not only the $\chi'$
maximum near 4.5 K but a shoulder $\sim 7 K$ as well. In addition
$\chi''$ has two maxima, at 4.5 K and at $\sim 7$ K. The differences
we observe in Fig.~\ref{fig:chifield3pl} with application of very
small fields, as well as the appearance of structure in the $T$-sweeps
when the field is carefully zeroed are extraordinary. We interpret
this sensitivity as an indication of enormous degeneracy, or near
degeneracy, of magnetic moment configurations in these crystals. We
observe that DC fields as small as 10 Oe have a significant effect,
decreasing the peak value of $\chi'$ by a factor of 3.5 and $\chi''$
by more than a factor of 30.  The reason for the marked difference
between the AC and DC magnetic susceptibilities that we noted above is
now quite clear; the DC magnetic susceptibility was measured at fields
between 50 and 10$^4$ Oe applied (DC) field, whereas $\chi'$ shows
dramatic changes between 0 and 200 Oe in the range $1.8 <T< 20 K$ for
all samples measured with AC susceptibility techniques at finite $H$.
Therefore, the AC susceptibility data represent a clearer impression
of the $H=0$ magnetic state of our Fe$_{1-x}$Co$_x$S$_2$ crystals. In
addition, the sensitivity to frequency and magnetic fields indicates a
complex energy landscape with nearly degenerate energy minima for the
magnetic moments. Our main conclusion from the broad temperature and
field dependence of $M$ and $C$ in sections
\ref{DCsusceptibilityandMagnetizationMeasurements},
\ref{ACSusceptibilityMeasurements}, and \ref{TheSpecificHeat}, that
clusters of magnetic moments form at $T>>T_c$, is consistent with the
conclusion of a spin-glass-like sensitivity to measurement $f$ and
applied magnetic fields over a broad $T$-range about $T_c$.

\begin{figure}[htb]
  \includegraphics[angle=90,width=3.5in,bb=70 90 540
  710,clip]{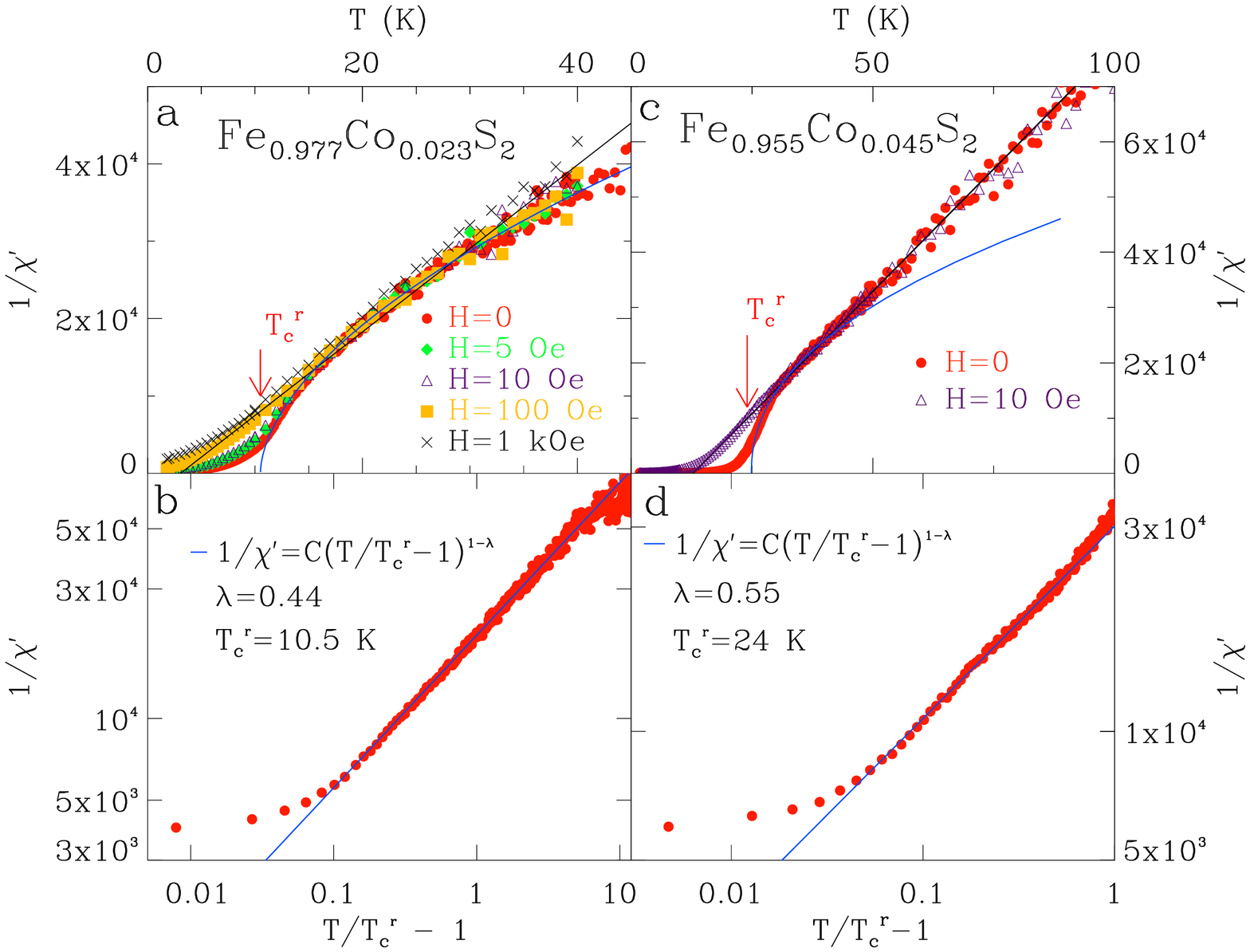}%
  \caption{\label{fig:ichivt} (Color online) Temperature dependence of
the inverse AC magnetic susceptibility. (a) The temperature, $T$,
dependence of the inverse of the real part of the AC magnetic
susceptibility, $1 / \chi'$ for $x=0.023$, at DC magnetic fields, $H$,
identified in the figure. The black line is the best fit of the
Curie-Weiss form to the $H=0$ data between 20 and 50 K with
$\Theta_W^{AC}=3.0$ K and a Curie constant corresponding to a
population of $1.5\times10^{21}$ cm$^{-3}$ spin 1/2 or 0.061 spin 1/2 per
Fe$_{0.977}$Co$_{0.023}$S$_2$ formula unit. The blue line is a fit of
Eq. \ref{eq:grifchi} in the text to the data with a best fit value of
$\lambda = 0.44 \pm 0.05$ and $T_c^r$ of $10.5 \pm 0.5$ K identified
in the figure by the red arrow. (b) $1/\chi'$ of our $x=0.023$ sample
as a function of reduced temperature, $T/T_c^r -1$, with $T_c^r$
determined from the best fit of Eq.\ \ref{eq:grifchi} to the data
above 11.5 K. The Blue line is the same as in frame a of the
figure. (c) $T$-dependence of $1 / \chi'$ for $x=0.045$, at $H$'s
identified in the figure. The black line is the best fit of the
Curie-Weiss form to the $H=0$ data between 25 and 100 K with
$\Theta_W^{AC}=12.8 \pm 0.5$ K and a Curie constant corresponding to a
population of $1.8\times10^{21}$ cm$^{-3}$ spin 1/2 or 0.072 spin 1/2 per
Fe$_{0.955}$Co$_{0.045}$S$_2$ formula unit. The blue line is a fit of
Eq. \ref{eq:grifchi} in the text to the data with a best fit value of
$\lambda = 0.55 \pm 0.05$ and $T_c^r$ of $24 \pm 0.5$ K identified in
the figure by the red arrow. (d) $1/\chi'$ of our $x=0.045$ sample as
a function of $T/T_c^r -1$ with $T_c^r$ determined from the best fit
of Eq.\ \ref{eq:grifchi} to the data above 26.5 K. The Blue line is
the same as in frame c of the figure.  Data taken with an excitation
field of 1 Oe at a frequency of 1 Hz.}
\end{figure}

   To aid in understanding the changes that occur to the magnetic
response of this system just above the ordering temperature, we have
compared $1/\chi'$ taken at small DC magnetic fields as a function of
$T$ to the Curie-Weiss form as shown in Fig.~\ref{fig:ichivt} for
$x=0.023$ and $x=0.045$. What is interesting here is that $1/\chi'$
falls below the Curie-Weiss behavior established at higher
temperatures. That is, $\chi'$ diverges more strongly than the
Curie-Weiss behavior for temperatures substantially above $T_c$. We
note that $1/\chi'$ evolves toward the Curie-Weiss form, $1 / \chi' =
(T - \Theta_W^{AC}) / CC$, with the application of small fields of
order 100 Oe for $x=0.023$ and only 10 Oe for $x=0.045$. The strong
increase of $\chi'$ above $T_c$ and $\Theta_W$ is evidence for short
ranged ferromagnetic order, or as we have noted above,
ferromagnetically aligned clusters of magnetic moment
formation\cite{ouyang}. We conclude from this form of $\chi'$ that the
ground state of this system is likely disordered ferromagnetic rather
than a spin-glass state. In the bottom half of Fig.~\ref{fig:ichivt}
we demonstrate that a power-law temperature dependent
form\cite{CastroNeto,salamon},
\begin{equation}1/\chi' \propto (T/T_c^r -1)^{1 -
\lambda}\label{eq:grifchi}\end{equation} with $\lambda \sim 1/2$,
describes $\chi'$ for 1.5 to 2 orders of magnitude in reduced
temperature above $T_c^r$.  Here $T_c^r$ is an indication of the
temperature scale where ferromagnetic clusters begin to form and is an
indication of the maximum interaction strength between magnetic
moments. In contrast, $\Theta_W^{AC}$ is an average, or mean field
interaction between magnetic moments. A compilation of $\lambda$ and
$T_c^r$ values from fits of Eq.~\ref{eq:grifchi} to $1/\chi'$ for a
small number of crystals with $x$ varying between 0.012 and 0.045 is
included in Fig.~\ref{fig:chidcpwl}. While $T_c^r$ tends to increase
with $x$, $\lambda$ takes on values between 0.2 and 0.4 for $x>x_c$.

\begin{figure}[htb]
  \includegraphics[angle=90,width=3.2in,bb=70 320 505
  690,clip]{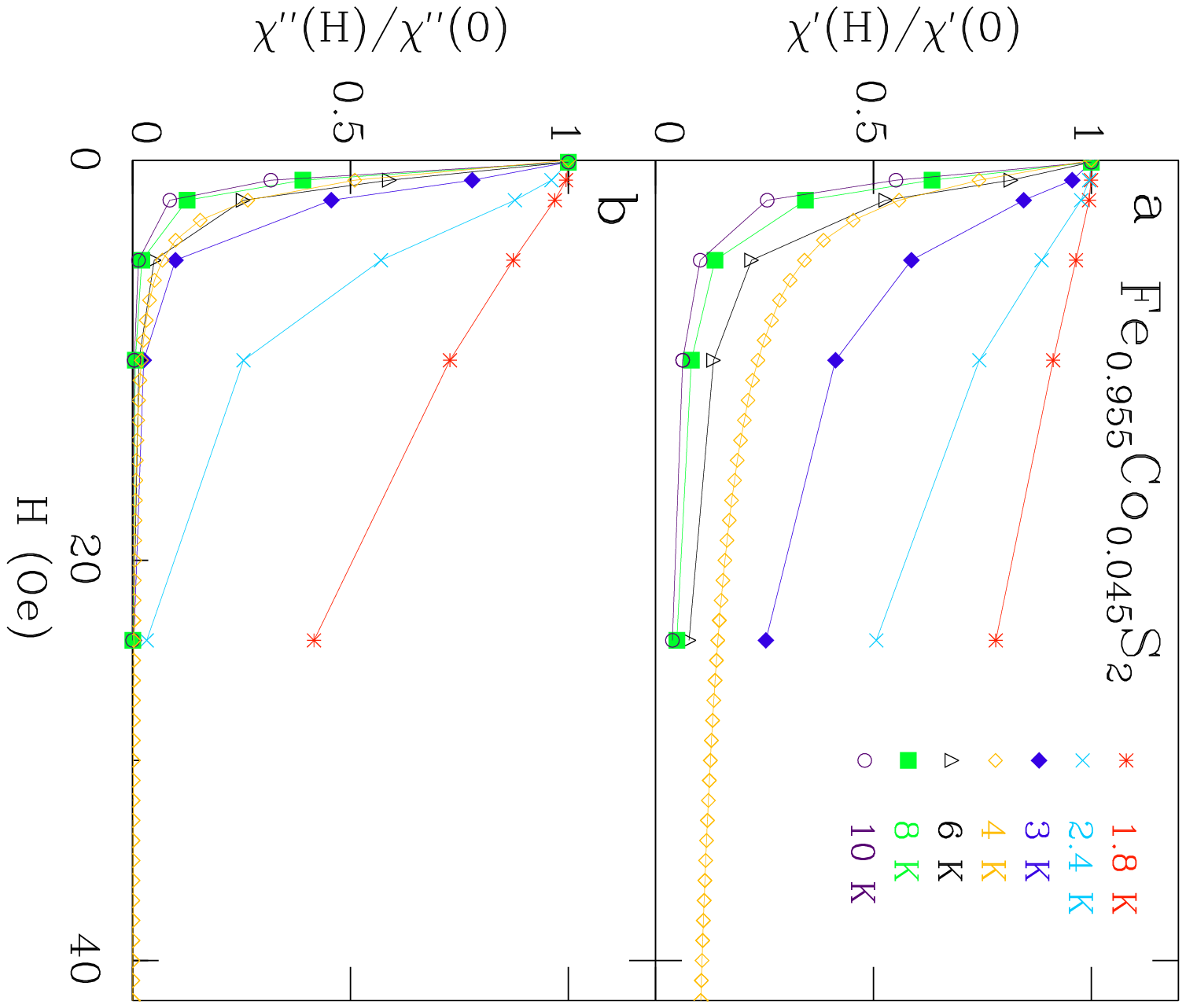}%
  \caption{\label{fig:chirivhfd1} (Color online) Magnetic field
dependence of the AC magnetic susceptibility. The magnetic field, $H$,
and temperature dependence of (a) the real part, $\chi'$, and
(b) imaginary part, $\chi''$, of the AC susceptibility, for the same
single crystal as in Figs.~\protect{\ref{fig:chifreqtemp}} and
\protect{\ref{fig:chifield3pl}} at temperatures identified in the
figure. Data taken with an excitation field of 1 Oe at a frequency of
1 Hz.}
\end{figure}

Beyond the enhanced frequency and magnetic field dependence seen in
Figs.~\ref{fig:chifreqtemp} and \ref{fig:chifield3pl}, there are
features in our data that indicate that the magnetic state of
Fe$_{1-x}$Co$_x$S$_2$ is far more interesting than that of a metallic
spin glass, particularly for temperatures above $T_c$.  For example,
in Fig.~\ref{fig:chirivhfd1} we plot the non-linear AC
susceptibility, $\chi'(H)$ and $\chi''(H)$ at temperatures near $T_c$
showing explicitly the changes that occur with $H$ at constant
$T$. Although this figure shows a broad, continuous, decrease of both
$\chi'$ and $\chi''$ with $H$ for $T<T_c$, it also demonstrates a very
sharp low field behavior for $T\ge T_c$. A mean-field analysis of the
non-linear susceptibility involves the assumption of analyticity
around $H=0$ such that a power series expansion in even powers,
$\chi'(H) = \chi_1 + \chi_3H^2 + \chi_5H^4 + ...$ can be employed to
describe $\chi'(H)$. The extraordinary field dependence of $\chi'$
seen in Fig.~\ref{fig:chirivhfd1} cannot be described accurately by
this form even over a reduced field range of only 10 Oe for
$T>T_c$. Thus the randomness due to the chemical substitutional
disorder has modified the field response of this system above $T_c$
such that it is no longer analytic near $H=0$.

This form of $\chi'(H)$ is similar to that found in (body-centered
tetragonal) LiHo$_{x}$Y$_{1-x}$F$_4$ where the magnetic moments
associated with the Ho ions strongly align along the c-axis and couple
to each other via the dipolar interaction\cite{wu,silevitch}. The
ground state is ferromagnetic for $x>0.2$ and a spin-glass for
$0.1<x\le0.2$\cite{wu,silevitch}. A magnetic field applied transverse
to the moments, $H_t$, causes a mixing of excited states with the
ground state doublet. This suppresses the ordering such that a quantum
critical point ($T=0$) has been accessed for a large range of $x$. For
$x<0.5$ Refs. \onlinecite{wu,silevitch,ancona} find that
$\chi'(H_t\rightarrow 0)$ and $\chi'(H_l\rightarrow 0)$, where $H_l$
is along the $c$-axis, both have a finite slope, $d\chi'/dH$, and thus
cannot be expressed in a perturbative expansion in $H$. The
interpretation was that the disorder induced field distribution
contains rare large amplitude random fields that prevent a description
of $\chi(H_t)$ in a mean field way as $H\rightarrow 0$. The authors
conclude that the non-analyticity at $H_t=0$ and $T>T_c$ is a
manifestation of Griffith singularities\cite{Griffiths}. Although
there are similarities between our system, Fe$_{1-x}$Co$_x$S$_2$, and
LiHo$_x$Y$_{1-x}$F$_4$, both are disordered ferromagnets that are
derived by chemical substitution from an insulating parent compound,
there are significant differences which are important to point
out. The most obvious is that Fe$_{1-x}$Co$_x$S$_2$ is a metal for $x
\ge 3\times10^{-4}$, while LiHo$_x$Y$_{1-x}$F$_4$ remains insulating
for all $x$. In addition, LiHo$_x$Y$_{1-x}$F$_4$ is strongly
Ising-like with the magnetic moments preferring to align along the
$c$-axis. In contrast, Fe$_{1-x}$Co$_x$S$_2$ is cubic and for $x=1$
has been shown to have a very small cubic anisotropy of its
magnetization so that it been described as
Heisenberg-like\cite{hiraka}. However, we find that both of these
disordered magnets contain rare large amplitude fluctuations that are
the hallmarks of Griffiths phase formation for $T>T_c$. Thus, not only
do we find evidence for such phases in samples on the verge of
magnetic ordering at $T=0$, but also discover evidence for similar
rare region effects at larger $x$, $x>x_c$ above the ordering
temperature.

\begin{figure}[htb]
  \includegraphics[angle=90,width=3.2in,bb=80 313 540
  715,clip]{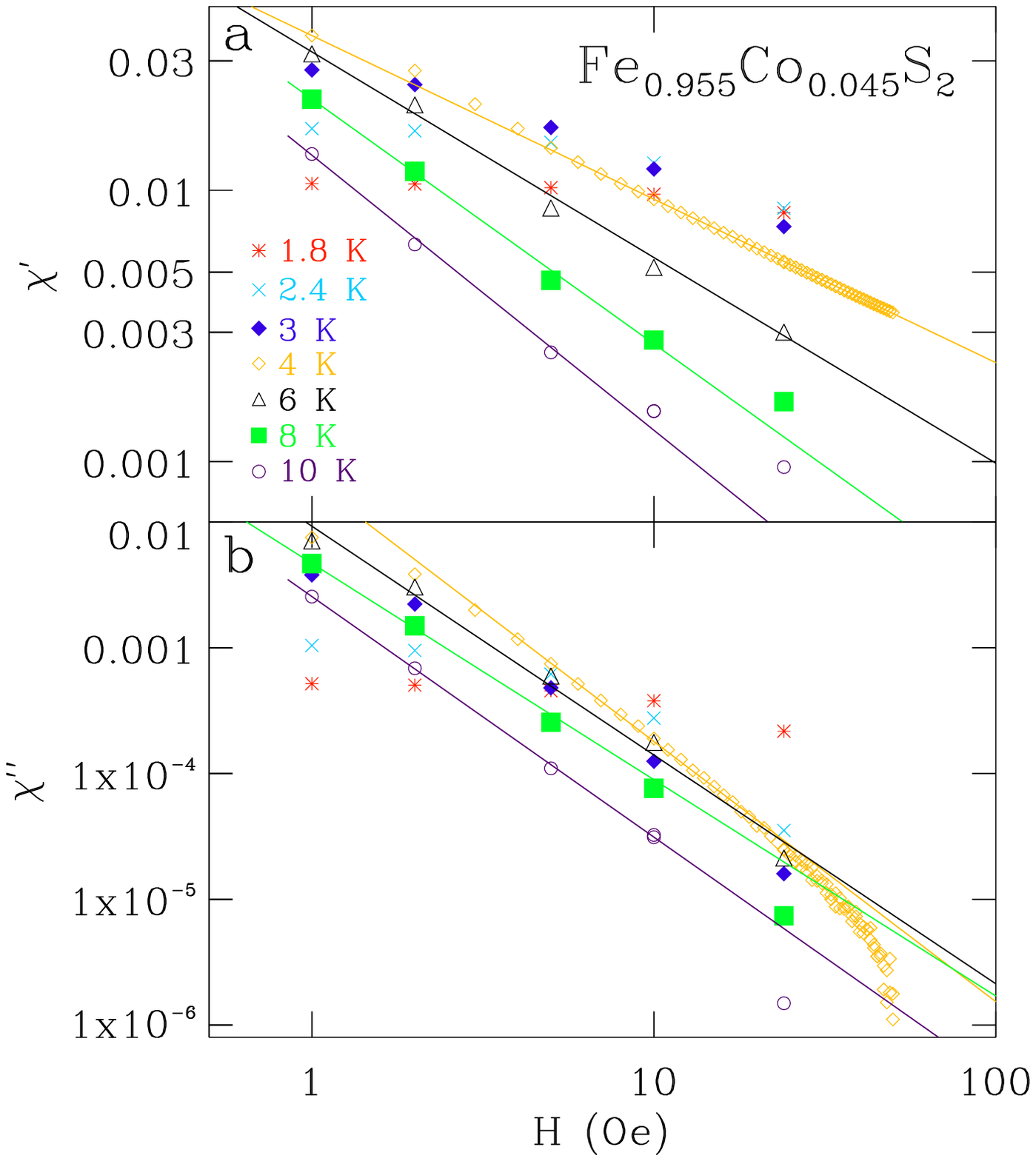}%
  \caption{\label{fig:chirivhfd} (Color online) Power law magnetic
field dependence of the AC susceptibility. The magnetic field, $H$
dependence of (a) the real part, $\chi'$, and (b) the imaginary part,
$\chi''$, of the AC magnetic susceptibility with logarithmic axis for
the same $x=0.045$ sample as in Figs.~\protect{\ref{fig:chifreqtemp}},
\protect{\ref{fig:chifield3pl}} and \protect{\ref{fig:chirivhfd1}} at
temperatures identified in the figure. The lines are power-law fits to
the data with best fit exponents of -0.6 (-2.1) at 4 K, -0.8 (-1.5) at
6 K, -0.9 (-1.7) at 8 K and -1.0 (-1.9) at 10 K for $\chi'$
($\chi''$). Data taken with an excitation field of 1 Oe at a frequency
of 1 Hz.}
\end{figure}

\begin{figure}[htb]
  \includegraphics[angle=90,width=3.2in,bb=80 320 540
  705,clip]{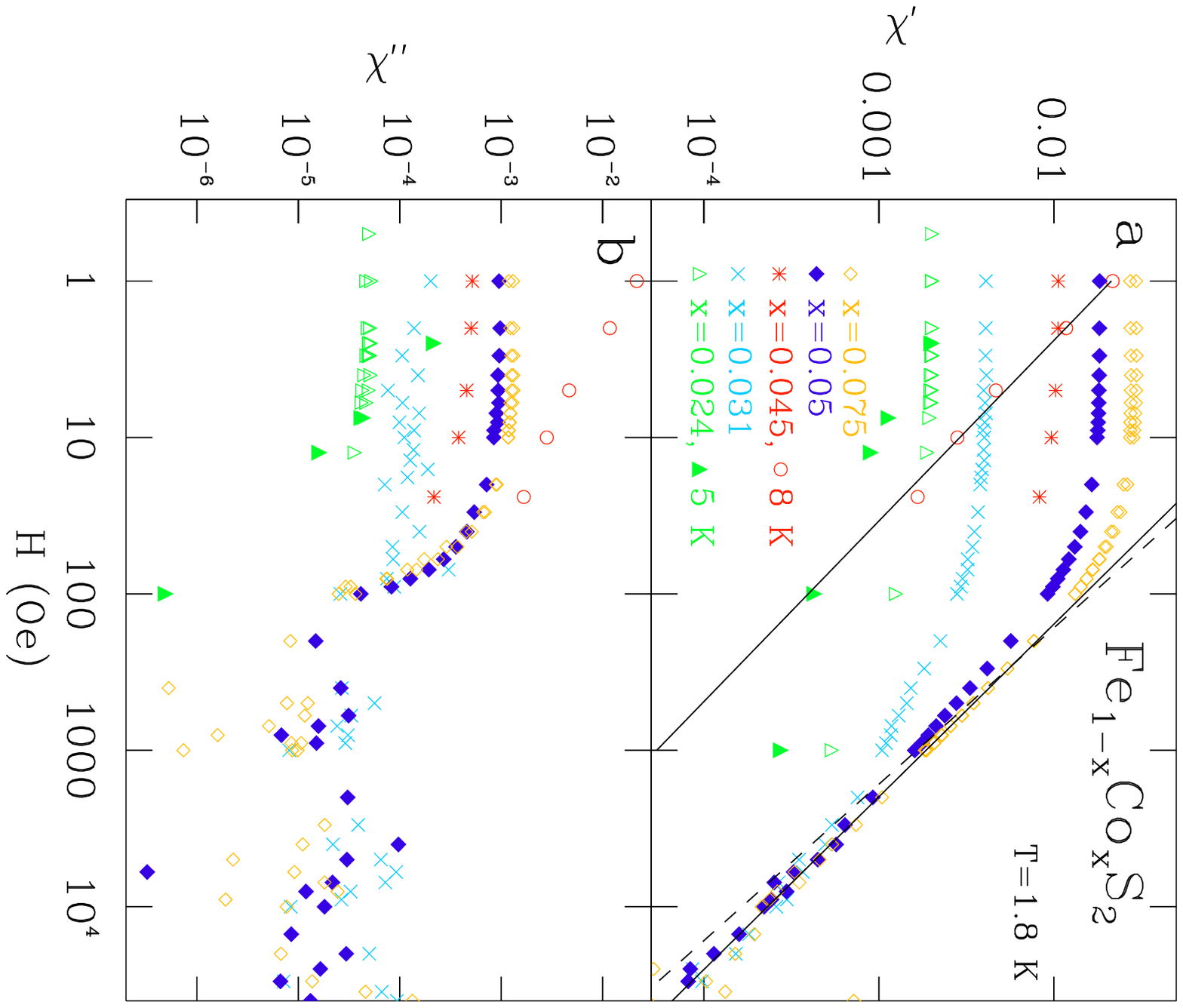}%
  \caption{\label{fig:chirivh} (Color online) Magnetic field
dependence of the AC susceptibility for various Co densities larger
than critical concentration. The magnetic field, $H$ dependence of (a)
the real part, $\chi'$, and (b) the imaginary part, $\chi''$, of the AC
magnetic susceptibility at $1.8$ K with logarithmic axis for
Fe$_{1-x}$Co$_x$S$_2$ for $x$'s identified in the figure. Data at one
$T>T_c$ are shown for the $x=0.024$ (5 K) and $x=0.045$ (8 K) crystals
to demonstrate the changes that occur to the low field behavior. The
solid lines are power-law fits to the data with best fit power-law of
-0.9 for the high field data at $T=1.8$ K as well as for the data for
$x=0.045$ at 8 K. Dashed line is a $H^{-1}$ dependence for
comparison. Data taken with an excitation field of 1 Oe at a frequency
of 1 Hz.}
\end{figure}

In contrast to mean-field forms, the magnetic field dependence of
$\chi'$ shown in Fig.~\ref{fig:chirivhfd1} can instead be well
described as a simple power-law in $H$ at $T>T_c$ as demonstrated in
Figs.~\ref{fig:chirivhfd} and \ref{fig:chirivh}. These figures
demonstrate that a small power-law form, $\chi' = \chi_0 H^{\alpha_M
-1}$, with $\alpha_M$ between 0 and 1 describes the data over at least
a decade and 1/2 in field for $T>T_c$, while for $T<T_c$ a much more
gentle field dependence is seen. The best fit $\alpha_M$ values have
been included in Fig.~\ref{fig:chidcpwl} for two samples where we have
sufficient data for an accurate determination of this parameter. In
addition, $\chi''$ can also be described in a power-law form with an
exponent between -1.3 and -2.0. For fields between 50 and 100 Oe
$\chi''$ is seen to undergo a much steeper decrease so that above
these fields it is consistent with zero being no larger than the
backgrounds in our measurement. The small power-law form for
$\chi'(H)$ at $T>T_c$ is consistent with the Griffiths phase
hypothesis made on the basis of the temperature dependence above.

Fig.~\ref{fig:chirivh} displays $\chi'$ and $\chi''$ at 1.8 K for 5
different samples with $x>x_c$ and demonstrates several features that
are common. First we note that for these Co concentrations $T_c > 1.8$
K and, as discussed above, $\chi'$ is not strongly field dependent
below about 100 Oe at this temperature. This is in contrast to the
data shown above $T_c$, a small subset of which is shown in the figure
for comparison.  For larger fields, $\chi'$ is seen to decrease as
$H^{-0.9\pm 0.1}$ while $\chi''$ is suppressed to within the
background level of zero above this same field scale.  It is
interesting to note that $\chi'$ above $10^3$ Oe is less dependent on
$x$ suggesting that here the magnetic energy level distribution is
much more independent of $x$.

Our presentation in this section has focused on the AC magnetic
susceptibility of our Fe$_{1-x}$Co$_x$S$_2$ crystals with $x>x_c$
which demonstrates that we have discovered several unusual
aspects. This includes a highly frequency dependent $\chi'$ and
$\chi''$ which encompasses a temperature range both above and below
$T_c$ that is much larger than in prototypical spin-glass systems. We
have found an extraordinary non-analytic field dependence to the AC
susceptibility near $H=0$ coupled with a small power law form to the
temperature dependence. These aspects are indicative of an
inhomogeneous magnetic system where rare ordered regions form at
temperatures above the long-range ordering, or perhaps a glass-like
freezing, temperature of the disordered system. These local
ferromagnetically ordered regions are well described by the Griffiths
phase phenomenology\cite{Griffiths,vojtarev,vladrev,CastroNeto} where
the non-universal power-law dependent properties are the result of
quantum mechanical tunneling of these rare regions to states with
different magnetizations.

\section{Discussion and Conclusions}
In this paper we have described a set of experiments exploring the
nucleation of a magnetic ground state in a carrier-doped nonmagnetic
insulator. For this investigation we chose the relatively simple
diamagnetic insulator FeS$_2$ as it allows for Co substitution for Fe
without the formation of second
phases\cite{Jarrett,bouchard,wang,ramesha,cheng}. In addition, end
member of the series CoS$_2$ is an itinerant ferromagnet with a Curie
temperature of 120 K\cite{bouchard}. We found that the Co substitution
yields a small number of itinerant charge carriers\cite{guoprb2} and
magnetic moments that are likely localized to the Co impurity sites. A
weakly coupled magnetic ground state, spin glass or disordered
ferromagnetic, developed for $x \ge 0.007 \pm 0.002$ as evidenced by a
peak in the temperature dependence of the magnetic susceptibility. As
$x$ is increased beyond $x_c$ the magnetic state becomes more robust
to temperature and magnetic field so that the FM state is more evident
at fields and temperatures where common tests can easily be performed.

In the paramagnetic regions of temperature and Co concentration, we
found fluctuating magnetic moments whose size was significantly larger
than that expected for spin-1/2 moments of individual Co dopants in an
FeS$_2$ background. This led us to the conclusion that clusters of
magnetic moments were forming upon cooling below 10 K for samples with
$x<x_c$ and at $T>T_c$ for sample with $x>x_c$. The formation of
clusters of average size of 3 to 5 $J=1/2$ magnetic moments was also
found to be consistent with the measured $C(T)/T$ and the entropy
derived from it. Because of the disorder and assumed random placement
of Co impurities throughout our samples, it is likely that a
distribution of cluster sizes results.  This distribution is likely to
have a very long tail for large cluster sizes so that there may be
rare regions of local magnetic order. Such a description fits in well
with the sensitivity to magnetic fields that we observed in the
susceptibility and magnetization of our crystals.

We posit that it is these rare regions of incipient order that cause
the unusual temperature and magnetic field dependencies that we
measured in the magnetic susceptibility and specific heat. The
consistent appearance of small power-laws in our analysis of both the
low temperature behavior of samples in proximity to $x_c$ and at
temperatures just above $T_c$ for samples with $x>x_c$ lead us to
conclude that Griffiths phase physics is the most likely mechanism
determining the static and dynamic properties of this
system\cite{vojtarev,vladrev}. In this scenario large rare regions of
order appear in disordered magnets at temperatures below the critical
temperature of the system without disorder. These ordered regions
which act as single large magnetic moments are subject to tunneling
events where the magnetic moment reverses. When a distribution of
cluster sizes is included, small power-law temperature dependences are
predicted. Details such as the anisotropies (Ising-like or
Heisenberg-like moments) as well as the mechanisms for dissipation are
thought to determine the observability conditions for these
effects\cite{vojtarev,vladrev}. Fe$_{1-x}$Co$_x$S$_2$ appears to be a
system where such effects may be observable over a wide temperature
range as the Kondo temperatures we measure are small, of order 1
K\cite{guo}, while the Weiss temperatures are typically much larger,
so that dissipation by way of the charge carriers may be relevant only
at the lowest temperatures.

It is no surprise that the disorder inherent to a doped semiconducting
system can lead to the formation of an inhomogeneous magnetic state at
low temperatures. What we find extraordinary are the unusual power-law
temperature and magnetic field dependencies of physical properties
that we measure along with the very sharp, non-analytic, field
dependence of the magnetic susceptibility near $H=0$. That a simple
model of disordered magnets\cite{Griffiths} which postulates the
existence of rare ordered regions can describe the broad features of
such a complicated system is both surprising and compelling.

\section{Acknowledgments}
We thank I. Vekhter and C. Capan for discussions.  JFD, DPY, and JYC
acknowledge support of the NSF under DMR084376, DMR0449022, and
DMR0756281.  We thank J. M. Honig for providing crystals of
V$_{0.99}$Ti$_{0.01}$O$_3$ for use as a manometer in our pressure
cells.


\end{document}